\documentclass[%
 aip,amsmath,amssymb,reprint,superscriptaddress
]{revtex4-1}

\usepackage{graphicx}
\usepackage{dcolumn}
\usepackage{bm}
\usepackage{enumerate}
\usepackage{eurosym}
\usepackage{amsfonts}
\usepackage{amsmath}
\usepackage{amsopn}
\usepackage{amssymb}
\usepackage{graphicx}
\usepackage{graphics}
\usepackage[colorlinks=true, allcolors=blue]{hyperref}
\usepackage{color}
\usepackage{verbatim}
\usepackage{mathrsfs}
\usepackage{hyperref}
\usepackage{comment}
\usepackage{braket}
\usepackage{graphicx}
\usepackage{dcolumn}

\makeatletter
\def\@email#1#2{%
 \endgroup
 \patchcmd{\titleblock@produce}
  {\frontmatter@RRAPformat}
  {\frontmatter@RRAPformat{\produce@RRAP{*#1\href{mailto:#2}{#2}}}\frontmatter@RRAPformat}
  {}{}
}%
\makeatother

\newcommand{\tcd}{\affiliation{School of Physics, CRANN and Rinn Quantum, Trinity College Dublin, College Green, Dublin 2, Ireland}}

\begin{document}

\preprint{AIP/123-QED}

\title{Finite-temperature Quantum Quasinormal Mode Theory for Spatiotemporally-resolved Dissipative Cavity Quantum Magnonics}
\author{Lars Meschede}
\tcd
\author{Daniel D. A. Clarke}
\tcd
\email{clarked5@tcd.ie}
\author{Ortwin Hess}
\tcd
\email{ortwin.hess@tcd.ie}

\date{\today}

\begin{abstract}
    Cavity quantum magnonics has ushered in an unprecedented era of hybrid light–matter interfaces in which microwave photons, collective spin excitations, and superconducting or spin qubits can be coupled within a single device architecture, facilitating a versatile platform for quantum information processing. Realizing their full potential, however, requires a theory that can treat non-reciprocal magnetic media, material dissipation and open-cavity radiation losses, as well as finite-temperature fluctuations rigorously and without significant approximation. The dissipative magnonic resonances of such systems are described as quasinormal modes (QNMs), yet a consistent quantization procedure for lossy gyrotropic resonators has remained elusive. In this work, we present a finite-temperature quantum QNM theory, based on macroscopic quantum electrodynamics for non-reciprocal media and complex coordinate transformations, that achieves QNM quantization for a wide class of spatially inhomogeneous, dissipative and dispersive, cavity magnonic systems. We demonstrate the predictive capabilities of our theory via applications to hybrid dielectric-magnonic resonators interacting with both spin and superconducting qubits, evidencing modal interference effects, a continuous transition from coherent to dissipative coupling, spatiotemporal quantum field dynamics, and magnon-mediated qubit synchronization, at both zero and finite temperatures. By addressing the intricacies of modal loss and non-reciprocity in a fully quantum theory of cavity magnonics, our approach enables the quantitative design of hybrid quantum devices that operate under realistic open-system and thermal conditions, without recourse to {\it ad hoc} quantization schemes.  
\end{abstract}

\maketitle

\section{Introduction}

Magnonics has emerged as a promising platform for the coherent manipulation and transport of information by means of spin waves, the collective excitations of ordered magnetic systems. Owing to their wave nature, magnons enable low-dissipation information processing and afford access to a rich variety of non-reciprocal\cite{wang2019nonreciprocity,kim2023nonreciprocity}, non-linear\cite{zheng2023tutorial}, non-hermitian\cite{kim2026non}, and topological\cite{wang2021topological, yu2024mag} phenomena. Advances in materials engineering and nanofabrication have substantially extended the capabilities of magnonic devices, motivating their exploration for both classical and quantum technologies, ranging from compact resonators to bulk and thin-film systems for propagation. Among available material platforms, yttrium iron garnet (Y$_3$Fe$_5$O$_{12}$, YIG) occupies a position of particular importance by virtue of its exceptionally low magnetic loss, long coherence times, high spin density that permits strong coupling to radio-frequency (RF) and microwave (MW) fields, and readily accessible nonlinear regime\cite{mallmann2013yttrium}.
These properties have already established YIG as a standard medium in commercial microwave systems, including filters and phase shifters\cite{levchenko2026review}.
At the quantum level, magnons are the quanta of the collective spin waves and are described as bosonic quasiparticles. Their low energy consumption, on-chip integrability, nanoscale wavelengths, frequency tunability under an applied bias field, and natural interface with MW circuitry and other classes of quantum excitation render them a promising resource for classical and quantum information processing and for integration into quantum computing architectures\cite{khitun2025magnonic,jiang2023integrating,yuan2022quantum,serha2026magnetic}.

A particularly rich vein of research is cavity magnonics, which addresses the interaction between magnons and confined MW photons. Hybrid cavity-magnonic systems have demonstrated both strong\cite{zhang2014strongly} and ultrastrong light-matter coupling\cite{goryachev2014high,bourhill2016ultrahigh}, thereby enabling the coherent exchange of excitations and the formation of magnon-polaritons. Such systems offer substantial potential for quantum memories\cite{zhang2015magnon}, optical-to-MW transduction via magneto-optical effects (cavity optomagnonics)\cite{hisatomi2016bidirectional,lachance2019hybrid,wang2022recent}, magnon-photon-phonon coupling mediated by magnetostrictive forces (cavity magnomechanics)\cite{shen2022coherent,zhang2016cavity}, the engineering of synthetic gauge fields\cite{gardin2024engineering}, and even more exotic applications such as axionic dark matter detectors\cite{barbieri1989axion,barbieri2017searching}.
At the quantum level, optical-to-MW transducers are expected to play an increasingly important role in quantum networks, since the majority of leading quantum-computing platforms operate at MW frequencies, whereas optical photons propagating in fiber are much better suited to long-distance quantum communication. In addition to the coherent interaction, dissipative magnon-photon~\cite{wang2020dissipative} and magnon-magnon~\cite{nair2022cavity} coupling, arising from the cavity Lenz effect~\cite{harder2018level} and from coupling to a common radiative reservoir~\cite{yao2019microscopic} can be observed, leading to level attraction and possibly exceptional points~\cite{liu2019observation,zhang2017observation} (coalescing eigenmodes) rather than the level repulsion characteristic of coherent coupling; the relative contributions of coherent and dissipative coupling can even be tuned~\cite{harder2018level,boventer2019steering}.

An instrumental route towards quantum applications is the interfacing of cavity magnonic systems with quantum bits (qubits) required for quantum information processing and computing. The arguably most promising qubit platforms, namely spin qubits (SQs) and ensembles thereof\cite{burkard2023semiconductor} (\textit{e.g.}, nitrogen-vacancy centers in diamond\cite{pezzagna2021quantum}, rare-earth ion-doped crystals\cite{bertaina2007rare}, defects in silicon carbide\cite{capan2025defects}) and superconducting (SC) qubits\cite{krantz2019quantum} (\textit{e.g.}, flux qubits\cite{dmitriev2021perspective} and transmon qubits\cite{roth2021introduction}), have already been interfaced with (cavity) magnonic systems in a variety of configurations.
Direct coupling of SQs to magnonic modes has been investigated both theoretically and experimentally, demonstrating coherent strong coupling for ferromagnets and ferrimagnets\cite{neuman2020nanomagnonic,candido2021predicted}, as well as for van der Waals materials~\cite{garcia2026strong}, long-distance coupling between SQs mediated by a ferromagnetic coupler~\cite{trifunovic2013long}, simultaneous magnon-photon lasing\cite{ribeiro2025dual}, and single-spin magnetometry at room temperature\cite{trifunovic2014high}. 

Experimentally, SC qubits have so far been coupled to magnons only indirectly, via a MW cavity mode\cite{tabuchi2015coherent,tabuchi2016quantum}.
This arrangement has enabled strong coherent magnon–qubit coupling, the generation of non-classical magnon states\cite{fan2023nonclassical}, dispersive coupling and single-quantum magnon sensing\cite{wolski2020dissipation,rani2025high,lachance2017resolving}, as well as both Kerr and cross-Kerr non-linearities\cite{lachance2017resolving,hu2011cross,holland2015single}.
Direct coupling of SC qubits to magnonic modes through the stray field has not yet been realized. Recent theoretical studies have nevertheless examined the direct coupling of transmon qubits to YIG microresonators, establishing the feasibility of generating non-classical magnonic states\cite{kounalakis2022analog,kounalakis2023engineering,ren2025preparing,jin2023magnon} and of implementing quantum gates\cite{dols2024magnon}. To the best of our knowledge, no experiment has demonstrated the coupling of a flux qubit (FQ) to a magnonic mode. Theoretical work has, however, explored coherent transfer between a FQ and a SQ via a YIG nanoresonator to which both are directly coupled\cite{lai2018single} (with the FQ placed outside the magnetic bias field), as well as the interaction of a FQ with a small, in-plane-magnetized YIG microsphere situated inside the FQ loop\cite{hou2024robust,liu2026generation}.   

Recent experimental advances further motivate the investigation of directly coupled magnon–qubit systems. FQs that operate at the flux-insensitive sweet spot with zero external bias flux have been demonstrated\cite{kim2024superconducting}, thereby simplifying on-chip integration and the interfacing with resonators by eliminating the need for an additional flux-bias line. In parallel, a fluxonium qubit capable of operation in the presence of in-plane magnetic fields in the tesla range has been realized\cite{gunzler2025spin}, an important step toward the experimental feasibility of direct FQ–magnon coupling. It is also possible to interface SC qubits and SQs (or SQ ensembles) directly so as to enable coherent exchange, as has been shown for both ensembles\cite{zhu2011coherent,twamley2010superconducting,marcos2010coupling} and single spins\cite{chang2025strong}; such interfaces offer a potential quantum bus for coupling SQs to other systems, including magnonic resonators. Hybrid SQ-SC-qubit architectures could therefore become valuable components of quantum processors, combining the long coherence times of SQs for storage with the stronger couplings characteristic of SC qubits for computation.

Given the diverse phenomenology and underlying interaction mechanisms in cavity quantum magnonics and hybrid architectures based upon it, an urgent demand for a rigorous and unified methodology has emerged, which would allow for a quantitative, spatiotemporally-resolved description of the quantum electromagnetic field dynamics and its interplay with various qubit realizations. Classically, the magnonic modes of ferro- and ferrimagnetic resonators were first described via a magnetostatic approach, namely the coupled Maxwell and Landau-Lifshitz (LL) equations under the approximation of negligible exchange interaction, Gilbert damping, propagation and radiation losses, giving rise to the well-known Walker modes with real frequencies\cite{walker1957magnetostatic,walker1958resonant,fletcher1959ferrimagnetic,solt1960magnetic}. The simplest such mode, in which all involved spins precess uniformly, is the Kittel mode and is typically observed in ferromagnetic resonance experiments\cite{kittel1948theory}, whereas higher-order non-uniform modes can be observed in inhomogeneous bias fields\cite{white1956multiple,goryachev2014high}. Still, this approach does not take into account the influence of a magnetic body's dielectric environment, as needed for the description of magnon-photon hybrid systems. A more general classical approach was used in Ref.~\cite{zare2015magnetic}, in which the electromagnetic scattering on cavity magnonic system was studied, based on generalized Mie theory \cite{lin2004electromagnetic,geng2004mie}. In this case, the Landau-Lifshitz-Gilbert (LLG) equation was solved, in which the exchange interaction was still neglected (valid for resonator sizes much larger than the exchange length) but where Gilbert damping was taken into account. Since the full Maxwell's equations were treated, propagation effects were also included. However, studying scattering alone may not be sufficient to gain a comprehensive understanding of cavity magnonic systems, since, for example, spectral splitting is only a necessary but not a sufficient condition for strong coherent coupling. A similar approach was used in Refs.~\cite{krupka2016electrodynamic,krupka2017ferromagnetic,krupka2019electrodynamic}, where the complex quasinormal modes (QNMs) for a bare and a shielded YIG sphere were calculated using an effective scalar permeability. In Ref.~\cite{macedo2020electromagnetic}, the magnon-photon coupling of a YIG sphere within a closed MW resonator was investigated using a perturbative approach to calculate the modified frequencies, assuming that the YIG sphere was significantly smaller than the photonic resonator. The quantum mechanical treatment usually starts from the full Hamiltonian of an interacting spin system, which can be reduced to a Hamiltonian of bosonic modes to describe collective excitations via the Holstein-Primakoff transformation\cite{holstein1940field}. In principle however, this transformation can only be used for infinite spin systems to derive a Hamiltonian and dispersion for the continuous-wave magnon modes. For compact resonators, the discrete modal frequencies are usually obtained from different approaches, like the magnetostatic theory, and are then used to construct a Hamiltonian with discrete bosonic modes\cite{tabuchi2016quantum,bittencourt2023magnomechanical}, with the losses added phenomenologically. The coupling to a photonic resonator mode is quantified via either an overlap integral of the magnetization arising from the magnonic mode and the magnetic field of the photonic mode\cite{soykal2010strong}, or an experimentally determined parameter\cite{bittencourt2023magnomechanical}. In such cases though, a systematic, {\textit a prior} prediction of the ratio between coherent and dissipative coupling is not possible.

A natural framework in which to describe magnonic modes and their coupling with the modes of photonic structures and qubits from first principles is the QNM theory~\cite{lai1990time,leung1994completeness,leung1994perturbation,leung1996dielectric,lee1999dyadic,kristensen2014modes,yan2018rigorous,lalanne2018light,sauvan2022norm}. Therein, the electromagnetic resonances of an open system correspond to the time-harmonic solutions of Maxwell's equations subject to suitable outgoing-wave conditions, being characterized by complex frequencies. The non-conservative nature of lossy cavity systems has raised a variety of non-trivial questions regarding a proper, mathematically rigorous formulation of normalization and orthogonality, alongside the issue of whether or not the QNMs truly form a complete set, and thus allow mode expansions for arbitrary electromagnetic field distributions.
In spite of the challenges and ambiguities, classical QNM methodology has undergone significant advances in the last decade, including improved understanding of the formal aspects of the theory~\cite{sauvan2022norm},
techniques for computing the QNMs of arbitrary resonators~\cite{bai2013efficient,sauvan2013theory,ge2014time,ge2014quasinormal,yan2018rigorous,lalanne2018light} and the release of general-purpose software~\cite{wu2023man}, as well as successful applications to the calculation of generalized effective mode volumes~\cite{kristensen2012mode,sauvan2013theory,kristensen2014modes}, electromagnetic near- and far-field characteristics~\cite{kongsuwan2020qnm,ren2020nfft,bedingfield2023multi}, Green's functions~\cite{lee1999dyadic,ge2014quasinormal}, enhanced spontaneous emission rates~\cite{sauvan2013theory,ge2014quasinormal,dezfouli2017modal,ren2024classical}, the semi-classical treatment of resonator-QE systems in the weak coupling regime~\cite{yang2015analytical} and elastic Purcell factors in optomechanical systems~\cite{elsayed2020elastic}. The purely classical electromagnetic QNM theory has also been transcended by the emergence of quantization protocols\cite{franke2019quantization,franke2020fluctuation,franke2020nonlinear,meschede2026quantum}, although to date, numerical studies have largely been confined to dissipative cavity quantum electrodynamics in nano-optical dielectric and plasmonic platforms. 

Recently, we introduced a unified theoretical framework, based on macroscopic quantum electrodynamics (MQED)\cite{scheel1998qed,dung1998mqed,gruner1996mqed,raabe2007unified,scheel2008macroscopic,raabe2008qed,philbin2010canonical,buhmann2012macroscopic} and complex coordinate transformations, that achieves QNM quantization for a wide class of three-dimensional (3D), spatially inhomogeneous, dissipative and dispersive, linear, magnetodielectric resonators\cite{meschede2026quantum}. The complex coordinate transformations equivalently convert the radiative losses into non-radiative material dissipation, and via a suitable transformation that reflects all the losses of the resonator, we define creation and annihilation operators that allow the construction of modal Fock states for the joint excitations of electric- and magnetic-field-dressed matter, providing the foundations for a quantum dynamic theory via
quantum Langevin and master equations~\cite{franke2019quantization}. Whilst our numerical demonstrations concerned Purcell-enhanced spontaneous emission rates and single-emitter strong coupling in open photonic resonators, our formalism bears the notable and, heretofore, unprecedented advantage among quantum QNM theories of accommodating both arbitrary permittivity and permeability distributions. As such, our approach addresses the pressing need for a quantized treatment of the lossy resonances of magnonic cavities and their interaction with spin and SC qubits without recourse to {\it ad hoc} quantization schemes. Moreover, our theory admits an important generalization to finite temperatures, offering an account of thermal corrections that are expected to be particularly significant in the RF and MW frequency ranges characteristic of experiments in cavity quantum magnonics.   

In this article, we present a finite-temperature quantum QNM theory for arbitrary, 3D, dissipative and dispersive,
cavity magnonic systems. Commencing from the classical QNM theory and MQED for non-reciprocal materials\cite{wu2021qnmnonreciprocal_resonators}, we generalize our PML-based QNM quantization approach of Ref.~\cite{meschede2026quantum} for lossy magnetic resonators in general dielectric environments, enabling the prediction of all magnon-photon couplings and hybridizations from first principles, revealing the quantitative limitations of purely phenomenological quantization protocols, and offering unprecedented insight into the quantum dynamics of magnon-qubit interactions in hybrid quantum architectures. Our methodology affords a truly 4D (spatially and temporally resolved) characterization of quantum magnonic systems and their key figures of merit for device design, optimization and application in the Quantum 2.0 era.

\section{\label{Sec:Classic_QNMs} Classical QNM Theory for Non-reciprocal Magnetic Materials}

In this section, we discuss the classical electromagnetic QNM theory for non-reciprocal media based on PML regularization. The construction follows Refs.~\cite{sauvan2013theory, vial2014quasimodal,yan2018rigorous,wu2021qnmnonreciprocal_resonators} and is reviewed in Ref.~\cite{lalanne2018light}. In particular, we focus on magnetic materials like YIG. Whilst a numerical solver for such problems is publicly available\cite{wu2021nanoscale,wu2021qnmnonreciprocal_resonators,sauvan2022norm}, utilizing an auxiliary-field formulation to treat gyrotropic resonators with dispersion, not all important derivations are. We hope to fill this gap here, by reviewing the formalism, generalizing it to include anisotropy and exchange interactions, and collecting the important formulae needed for the QNM quantization scheme.

\subsection{Linearized Magnetization Dynamics}

The magnetization dynamics of ferro- and ferrimagnets is commonly described using the LLG equation of classical micromagnetism:
\begin{align}\label{eqn:LLG_general}
    \partial_t \bm{M}(\bm{r},t) = -\gamma \bm{M}\times \bm{H}_\mathrm{eff} +\frac{\alpha}{M_\mathrm{s}} \bm{M} \times \partial_t \bm{M} \ ,
\end{align}
in which $\gamma$ is the gyromagnetic ratio, $\alpha$ is the Gilbert damping parameter and $M_\mathrm{s}$ the saturation magnetization.
The effective magnetic field strength $\bm{H}_\mathrm{eff}$ typically comprises the static external bias field $\bm{H}_0$, the applied oscillating RF/MW field $\bm{h}$, the magnetocrystalline anisotropy field $\bm{H}_\mathrm{ani}$, the static demagnetizing field of the saturated body, and if retained, the exchange field $\bm{H}_\mathrm{ex} = J\nabla^{2} \bm{M}$.
In general, the magnetocrystalline anisotropy field depends on the symmetry properties of the crystal and on the orientation of the magnetization relative to the crystal axes; in this work, we shall assume that the external bias field is directed along a principal axis (namely, the easy/hard axis or orthogonal to it for uniaxial anisotropy, or a cube edge for cubic anisotropy), so that the contribution to the linearized magnetization dynamics below assumes the form
$\bm{H}_{\mathrm{ani}} = -\bm{D}\cdot\bm{M}$, with $\bm{D}=\mathrm{diag}(D_x,D_y,D_z)$ and where the constants $D_i$ are determined by the functional derivatives of the anisotropy energy density~\cite{schneider1972effect,miltat2007numerical}. For an ellipsoidal body, the static demagnetizing field is uniform, $\bm{H}_{d}=-\bm{N}\cdot\bm{M}_{s}$, with $\bm{N}=\mathrm{diag}(N_x,N_y,N_z)$ satisfying $N_{x} + N_{y} + N_{z} = 1$~\cite{osborn1945demagnetizing}.
It contributes only to a constant shift of the bias field that enters the linearized dynamics. We fix the direction of the saturation magnetization along $\bm{e}_z$ and assume a weak oscillating field $\bm{h}$ that produces a small deviation from saturation, $\bm{M} = {M}_\mathrm{s}\bm{e}_z + \bm{m}$, with $\vert\bm{m}\vert \ll {M}_\mathrm{s}$. Adding the static demagnetizing field $\bm{H}_\mathrm{d} = - N_z M_\mathrm{s} \bm{e}_z$, linearizing, and performing a Fourier transform with respect to both the space and time variables then yields
\begin{align}\label{eqn:LLG_linearized}
    \mathrm{i}\omega \bm{m} = \bm{e}_z \times \bigg[ \gamma M_\mathrm{s} \bm{h} &- (\gamma H_0 - \mathrm{i}\omega \alpha - \gamma M_\mathrm{s} (D_z + N_z) \notag \\
    &+ \gamma M_\mathrm{s}\bm{D} + \gamma M_\mathrm{s} J k^2   ) \bm{m} \bigg]  \ .
\end{align}
Note that the dynamic demagnetizing field, $-\bm{N}\cdot\bm{m}$, is not added to $\bm{H}_{\mathrm{eff}}$, since it is generated automatically when the effective permeability derived here is introduced into the full Maxwell’s equations. Indeed, from the above equation, we can express the magnetization in the form $\bm{m} = \chi_\mathrm{m} \bm{h}$, and thus identify the effective permeability tensor $\mu = 1 + \chi_\mathrm{m}$ as
\begin{align}\label{eqn:mu_tensor_general}
    \mu(\omega, \bm{k}) = \begin{pmatrix}
        1 + \chi_{xx}(\omega, k) & -\mathrm{i}\kappa_r(\omega, k) & 0 \\
        \mathrm{i}\kappa_r(\omega, k) & 1 + \chi_{yy}(\omega, k) & 0 \\
        0 & 0 & 1
    \end{pmatrix} \ , 
\end{align}
where
\begin{align}
    \kappa_r = \frac{\omega \omega_\mathrm{M}}{ (\omega_\mathrm{H} + A_x \omega_\mathrm{M} - \mathrm{i}\alpha\omega)  (\omega_\mathrm{H} + A_y \omega_\mathrm{M} - \mathrm{i}\alpha\omega) - \omega^2} \ , \\
     \chi_{xx} = \frac{\omega_\mathrm{M} ( \omega_\mathrm{H} + A_y \omega_\mathrm{M} - \mathrm{i} \alpha\omega)}{ (\omega_\mathrm{H} + A_x \omega_\mathrm{M} - \mathrm{i}\alpha\omega)  (\omega_\mathrm{H} + A_y \omega_\mathrm{M} - \mathrm{i}\alpha\omega) - \omega^2} \ , \\
     \chi_{yy} = \frac{\omega_\mathrm{M} ( \omega_\mathrm{H} + A_x \omega_\mathrm{M} - \mathrm{i} \alpha\omega)}{ (\omega_\mathrm{H} + A_x \omega_\mathrm{M} - \mathrm{i}\alpha\omega)  (\omega_\mathrm{H} + A_y \omega_\mathrm{M} - \mathrm{i}\alpha\omega) - \omega^2}\ . 
\end{align}
Here, $\omega_\mathrm{M} = \gamma M_\mathrm{s}$, $\omega_\mathrm{H} = \gamma (H_0 - N_z M_\mathrm{s})$ and $A_i = \Delta D_{i} + J k^2$, with $\Delta D_{i} = D_i - D_z$. The $k$-dependence caused by the exchange interaction implies spatial dispersion (\textit{i.e.}, non-locality). This means that the constitutive relation for the magnetic field is given by a spatial convolution rather than a simple proportionality:
\begin{align}\label{eqn:mu_spatial_convolution}
    \bm{B}(\bm{r},\omega) &= [\mu(\omega) \star \bm{H}(\omega)](\bm{r}) = \int\mathrm{d}^3r'\, \mu(\bm{r},\bm{r}',\omega) \bm{H}(\bm{r}',\omega) \ ,
\end{align}
with the spatial inverse Fourier transform 
\begin{align}
    \mu(\bm{r},\bm{r}',\omega) = \frac{1}{(2\pi)^3}\int\mathrm{d}^3 k\, \mu(\bm{k},\omega) \mathrm{e}^{\mathrm{i}\bm{k}\cdot(\bm{r}-\bm{r}')} \  .
\end{align}
While this expression is valid for infinite magnetic media only, the real-space form of the non-local permeability for finite resonators might be difficult to determine analytically. The linearization of the QNM eigenvalue problem discussed in Sect.~\ref{subsec:left_right_QNMs} also provides a method to calculate it numerically.

It is noteworthy that the QNM formalism has already been used to describe spatial non-locality in plasmonic systems\cite{kamandar2017nonlocal} by means of the hydrodynamic Drude model, in which the metal response is treated via a wavevector-dependent permittivity\cite{mcmahon2009nonlocal}. There, an auxiliary field was introduced to capture the non-local response; it is essentially analogous to the auxiliary field used to linearize the QNM eigenproblem in the case of spatially local, but temporally non-local systems in standard QNM solvers\cite{yan2018rigorous}, with the well-known $\beta^{2}k^{2}$ pressure contribution restored as a spatial derivative after the inverse Fourier transform. As we now introduce similar auxiliary fields to linearize the present eigenproblem for a frequency-dependent permeability, the exchange interaction can be included in the same way, where the $Jk^{2}$ contribution in $\mu(\omega, \bm{k})$ is the magnetic counterpart of $\beta^{2}k^{2}$. 

\subsection{Left and Right QNMs and Auxiliary-field Formulation \label{subsec:left_right_QNMs}}

The QNMs are generally defined as time-harmonic solutions of the source-free Maxwell's equation with outgoing-wave conditions imposed. In the presence of non-reciprocal media, the frequency-domain eigenvalue problem lacks self-adjointness, so that one must distinguish right eigenmodes $(\bm{H}_n^{(\mathrm{R})}, \bm{E}_n^{(\mathrm{R})} )$ of the material parameters $(\mu, \varepsilon)$ from left eigenmodes $(\bm{H}_n^{(\mathrm{L})}, \bm{E}_n^{(\mathrm{L})})$ of the corresponding transposed ones $(\mu^{\mathrm{T}}, \varepsilon^{\mathrm{T}})$:
\begin{align}\label{classicalQNM}
    \begin{pmatrix}
    0 & -\mathrm{i}\left({\mu}^{-1}\right)^{(\mathrm{T})}\nabla\times \\
    \mathrm{i}\left({\varepsilon}^{-1}\right)^{(\mathrm{T})}\nabla\times & 0 \\
    \end{pmatrix}
    \begin{pmatrix}
    \bm{H}_n^{(\mathrm{L/R})} \\
    \bm{E}_n^{(\mathrm{L/R})}
    \end{pmatrix} \notag \\
    = {\tilde \omega}_n
    \begin{pmatrix}
    \bm{H}_n^{(\mathrm{L/R})} \\
    \bm{E}_n^{(\mathrm{L/R})}
    \end{pmatrix}\ .
\end{align}
The left and right QNMs share the same eigenvalues but not the same eigenvectors in the case of non-reciprocal materials. For proper normalization, both the left and right QNMs are required\cite{wu2021qnmnonreciprocal_resonators, wu2021nanoscale, sauvan2022norm}.

Reflecting the non-hermitian nature of the system, stemming from both radiation and material losses, the eigenvalues ${\tilde \omega}_n = \omega_n - \mathrm{i}\gamma_n$ are complex, which causes the QNMs to exponentially diverge in space for $|\bm{r}| \rightarrow \infty$. By introducing perfectly matched layers (PMLs), described by transformed material parameters $(\tilde{\mu}, \tilde{\varepsilon})$, the QNMs become exponentially damped inside the PML region and the infinite domain can thus be truncated to a finite computational one $\Omega_\mathrm{PML}$ with perfect electrical/magnetic conductor boundary conditions\cite{yan2018rigorous, sauvan2022norm, meschede2026quantum}. Due to the truncation, the continuum contribution is mapped to a discrete set of PML modes that complement the QNMs. In what follows, all of these modes will be treated on equal footing, and we will simply refer to them as QNMs or modes.

The frequency-dependence of the permeability renders the eigenvalue problem in Eq.~\eqref{classicalQNM} non-linear. Nevertheless, the introduction of suitably defined auxiliary fields facilitates its linearization, enabling both practical numerical computations and analytical derivations. For general non-reciprocal media, we define auxiliary magnetization and current fields on the resonator domain $\Omega_\mathrm{res}$ in the form 
\begin{align}\label{eqn:auxiliary_fields}
    \bm{M}^{(\mathrm{R})} = \frac{1}{\mathrm{i}\omega} \left(\mu^{-1} - \mu_\infty^{-1} \right) \nabla \times \mathbf{E}^{(\mathrm{R})} \ , \  \bm{J}^{(\mathrm{R})} = - \mathrm{i}\omega \bm{M}^{(\mathrm{R})} \ , \notag \\
    \bm{M}^{(\mathrm{L})} = \frac{1}{\mathrm{i}\omega} \left(\mu^{-\mathrm{T}} - \mu_\infty^{-1} \right) \nabla \times \mathbf{E}^{(\mathrm{L})} \ , \  \bm{J}^{(\mathrm{L})} = - \mathrm{i}\omega \bm{M}^{(\mathrm{L})} \ . 
\end{align}
For the right QNMs, we can write Eq.~\eqref{eqn:auxiliary_fields} as $\mathrm{i} (\mu^{-1} - \mu_\infty^{-1})^{-1} \bm{M}^{(\mathrm{R})} = \nabla\times\mathbf{E}^{(\mathrm{R})}$ and derive the following expression for the $xy$-subblock,
\begin{align}
    \left(\mu^{-1} - \mu_\infty^{-1}\right)^{-1} = -\frac{\mu_\infty}{\omega_\mathrm{M}} \left( (\omega_\mathrm{H} +\omega_\mathrm{M}) \bm{I} +\omega_\mathrm{M} \bm{A} \right) \notag \\
    + \mathrm{i} \omega\frac{\mu_\infty}{\omega_\mathrm{M}} \bm{\alpha} \ , 
\end{align}
in which $\mathbf{I}$ is the $2\times2$ identity matrix, $\mathbf{A} = \mathrm{diag}(A_x, A_y)$ and $\bm{\alpha} = \begin{pmatrix}
    \alpha & -1 \\
    1 & \alpha
\end{pmatrix}$. This allows us to rewrite the equation as 
\begin{align}\label{eqn:aux_field_eq}
    \mathrm{i}\frac{\omega_\mathrm{M}}{\mu_\infty} \bm{\alpha}^{-1}\nabla\times\mathbf{E}^{(\mathrm{R})} + \frac{1}{\mathrm{i}} \bm{\alpha}^{-1}( (\omega_\mathrm{H} +\omega_\mathrm{M} )\bm{I} + \omega_\mathrm{M}\bm{A} ) \bm{J}^{(\mathrm{R})} \notag \\
    = \omega \bm{J}^{(\mathrm{R})} \ , 
\end{align}
which is now linear in the eigenvalue. By using the definition of $A_{x/y}$ and the inverse Fourier transform, we can write Eq.~\eqref{eqn:aux_field_eq} in the form
\begin{align}\label{eqn:aux_field_eq2}
    \mathrm{i}\frac{\omega_\mathrm{M}}{\mu_\infty} \bm{\alpha}^{-1}\nabla\times\mathbf{E}^{(\mathrm{R})} + \frac{1}{\mathrm{i}} \bm{\alpha}^{-1}( (\omega_\mathrm{H} +\omega_\mathrm{M} )\bm{I} + \omega_\mathrm{M}\Delta\bm{D} ) \bm{J}^{(\mathrm{R})}\notag \\
    -\frac{1}{\mathrm{i}} \bm{\alpha}^{-1} \omega_\mathrm{M} J \nabla^2  \bm{J}^{(\mathrm{R})}= \omega \bm{J}^{(\mathrm{R})} \ ,
\end{align}
which now also contains differential operators acting on $ \bm{J}^{(\mathrm{R})}$ stemming from the exchange interaction, similar to the additional terms in the hydrodynamic Drude model\cite{kamandar2017nonlocal}. Consequently, the linearized eigenvalue problem can be expressed as $\hat{H}^{(\mathrm{R})}\psi = \omega \psi$, with the augmented vector $(\bm{H}^{(\mathrm{R})}, \bm{E}^{(\mathrm{R})},\bm{M}^{(\mathrm{R})},\bm{J}^{(\mathrm{R})} )^\mathrm{T}$ and the effective frequency-independent Hamiltonian
\begin{align}\label{eqn:lin_ev_problem}
    \hat{H}^{(\mathrm{R})} = \begin{pmatrix}
        0 & \frac{1}{\mathrm{i}\mu_\infty} \nabla\times & 0 & \mathrm{i} \\
        \frac{\mathrm{i}}{\varepsilon}\nabla\times & 0 & 0 & 0 \\
        0 & 0 & 0 & \mathrm{i} \\
        \mathrm{i}\frac{\omega_\mathrm{M}}{\mu_\infty}\bm{\alpha}^{-1}\nabla\times & 0 & 0 & \hat{H}^{(\mathrm{R})}_{J J}
    \end{pmatrix}
\end{align} 
with $\hat{H}_{JJ}^{(\mathrm{R})} =  \frac{1}{\mathrm{i}} \bm{\alpha}^{-1}( (\omega_\mathrm{H} +\omega_\mathrm{M} -\omega_\mathrm{M} J \nabla^2)\bm{I} + \omega_\mathrm{M}\Delta\bm{D} )$. The same construction for the left QNMs with transposed matrix $\bm{\alpha}  ^{\mathrm{T}}$ defines $\hat{H}^{(\mathrm{L})}$. In the absence of the exchange interaction ($J = 0$), the equations for $\bm{M}^{(\mathrm{L/R})}$/$\bm{J}^{(\mathrm{L/R})}$ are of algebraic nature without any restrictions for their boundary values on $\partial\Omega_\mathrm{res}$. For $J\neq 0$, they become partial differential equations, which have to be solved with the additional exchange boundary condition $\partial_\mathrm{n} \bm{M}^{(\mathrm{L/R})} = 0$. In the presence of additional surface torques (for instance, N{\'e}el surface anisotropy), the more general Rado-Weertman boundary condition has to be imposed~\cite{rado1959spin,guslienko2005boundary}.

This linearization permits an efficient computational solution of the QNM and scattering problems without explicit knowledge of the non-local permeability in real space. However, for some applications it might be required, as for the QNM quantization in \S~\ref{Sec:Quantization_QNMs}. The frequency-dependent non-local part of the permeability can be reconstructed by solving the eigenvalue problem $\hat{H}_{JJ}^{(\mathrm{L/R})} \bm{u}_l^{(\mathrm{L/R})} = \lambda_l \bm{u}_l^{(\mathrm{L/R})}$ on the resonator domain $\Omega_\mathrm{res}$, similar to Ref.~\cite{stancil1980magnetostatic}. Eq.~\eqref{eqn:lin_ev_problem} then leads to the representation:
\begin{align}\label{eqn:mu_spatial_expansion}
    &\left(\mu^{-1}(\omega) - \mu_\infty^{-1} \right)(\bm{r},\bm{r}',\omega) \notag \\
    &= \mathrm{i}\frac{\omega_\mathrm{M}}{\mu_\infty} \sum_l \frac{\bm{u}_l^{(\mathrm{R})}(\bm{r})\otimes \bm{u}_l^{(\mathrm{L})}(\bm{r}')}{\lambda_l - \omega} \bm{\alpha}^{-1} \ ,
\end{align}
with the normalized eigenfunctions $\int_{\Omega_\mathrm{res}} \mathrm{d}^3r\, \bm{u}_l^{(\mathrm{L})}(\bm{r})\cdot \bm{u}_l^{(\mathrm{R})}(\bm{r}) = 1$.

\subsection{Inner Product, (Bi-)Orthogonality and Completeness}
 
To derive an inner product and the orthogonality relation for the QNMs, we define the matrix
\begin{align}
    \hat{A} = \begin{pmatrix}
        -\mu_\infty & 0 & 0 & 0 \\
        0& \varepsilon & 0 & 0 \\
        0 & 0 & -\mu_\infty ( ( \frac{\omega_\mathrm{H}}{\omega_\mathrm{M}} - J \nabla^2 )\bm{I} + \Delta\bm{D} ) & 0 \\
        0 & 0 & 0 & 0
    \end{pmatrix} \ .
\end{align}
Utilizing the divergence theorem, we can ascertain the relation
\begin{align}\label{eqn:Lorentz_reciprocity_theorem}
    \int_\Omega \mathrm{d}^3 r\, \psi^{(\mathrm{L})}_{n} \hat{A} \hat{H}^{(\mathrm{R})} \psi^{(\mathrm{R})}_{n'} = \tilde{\omega}_{n} \int_\Omega \mathrm{d}^3 r\, \psi^{(\mathrm{L})}_{n} \hat{A}  \psi^{(\mathrm{R})}_{n'} \notag \\
    + \frac{1}{\mathrm{i}} \int_{\partial\Omega} \mathrm{d}\bm{s}\cdot [ \bm{E}^{(\mathrm{L})}_{n} \times\bm{H}^{(\mathrm{R})}_{n'}- \bm{E}^{(\mathrm{R})}_{n'} \times \bm{H}^{(\mathrm{L})}_{n}    ] \ ,
\end{align}
for an arbitrary domain $\Omega$. Using $\int_\Omega \mathrm{d}^3 r\, \psi^{(\mathrm{L})}_{n} \hat{A} \hat{H}^{(\mathrm{R})} \psi^{(\mathrm{R})}_{n'} = \tilde{\omega}_{n'} \int_\Omega \mathrm{d}^3 r\, \psi^{(\mathrm{L})}_{n} \hat{A}  \psi^{(\mathrm{R})}_{n'}$ leads to the unconjugated (source-free) form of the Lorentz reciprocity theorem~\cite{yan2018rigorous,wu2021qnmnonreciprocal_resonators}. When integrating over the entire PML-truncated domain $\Omega_\mathrm{PML}$, the boundary integrals in Eq.~\eqref{eqn:Lorentz_reciprocity_theorem} vanish due to the imposed boundary conditions, resulting in $\left( \tilde{\omega}_{n} -  \tilde{\omega}_{n'} \right) \int_{\Omega_\mathrm{PML}} \mathrm{d}^3 r\, \psi^{(\mathrm{L})}_{n} \hat{A}  \psi^{(\mathrm{R})}_{n'}  = 0$. Consequently, the orthonormality condition for the QNMs is given by
\begin{align}\label{eqn:QNMs_inner_product}
    \langle \psi_n^{(\mathrm{L})} |\psi_{n'}^{(\mathrm{R})} \rangle_A = \int_{\Omega_\mathrm{PML}} \mathrm{d}^3 r\, \psi^{(\mathrm{L})}_{n} \hat{A}  \psi^{(\mathrm{R})}_{n'} = \delta_{n n'} \ . 
\end{align}
Since it can be written as $\langle \psi_n^{(\mathrm{L})} |\psi_{n'}^{(\mathrm{R})} \rangle_A = \langle \hat{A}\psi_n^{(\mathrm{L})*} |\psi_{n'}^{(\mathrm{R})} \rangle_{L^2}$, the modes $\{\psi_n^{(\mathrm{R})}\}$ and their dual counterparts $\{\hat{A}\psi_n^{(\mathrm{L})*}\}$ form a biorthogonal basis with respect to the standard (complex) $L^2$-inner product.
The induced QNM norm $\mathrm{QN}_n^2 = \langle \psi_n^{\mathrm{L}} |\psi_{n}^{\mathrm{R}} \rangle_A  $ admits two equivalent expressions, namely
\begin{align}\label{eqn:QNM_norm_1}
    \mathrm{QN}_n^2 =  \int&_{\Omega_\mathrm{PML}}\mathrm{d}^3r\, \Bigg[ \bm{E}^{(\mathrm{L})}_n\varepsilon \bm{E}^{(\mathrm{R})}_n - \bm{H}^{(\mathrm{L})}_n\mu_\infty \bm{H}^{(\mathrm{R})}_n \\
    &- \bm{M}^{(\mathrm{L})}_n \mu_\infty\left( \left(\frac{\omega_\mathrm{H}}{\omega_\mathrm{M}} - J \nabla^2 \right) \bm{I} + \Delta\bm{D}  \right) \bm{M}^{(\mathrm{R})}_n \Bigg] \notag \
\end{align}
and
\begin{align}
    \mathrm{QN}_n^2 =  \int_{\Omega_\mathrm{PML}}&\mathrm{d}^3r\, \Bigg[ \bm{E}^{(\mathrm{L})}_n\varepsilon \bm{E}^{(\mathrm{R})}_n - \bm{H}^{(\mathrm{L})}_n \partial_\omega(\omega \mu(\omega) ) \bm{H}^{(\mathrm{R})}_n  \Bigg] \notag \ .
\end{align}
Henceforth, the QNMs are understood to be normalized so that $\mathrm{QN}_n = 1$, unless otherwise stated.
In the presence of the exchange interaction, the last equation would involve the calculation of a convolution integral requiring the analytical form of the non-local permeability, whereas this is not the case for the linearized equation.
Completeness of the QNMs alone within the resonator (for trivial background geometries) and of the set comprising both QNMs and PML modes everywhere inside the unmapped domain enclosed by the PMLs is assumed, as in the established dielectric and plasmonic contexts\cite{yan2018rigorous}; whilst no rigorous proof has been formulated for gyrotropic and possibly spatially dispersive media, it is here justified {\it a posteriori} by the agreement of truncated QNM expansions with the analytic Green's function in \S~\ref{Sec:Numerical Examples}.

\subsection{Modal Expansions and Green's Function}

Assuming completeness of the QNMs and PML modes, one can solve the generalized scattering/radiation problem of the form
$\hat{H}^{(\mathrm{R})} \psi^{(\mathrm{R})} = \omega \psi^{(\mathrm{R})} + S(\omega)$, where $S(\omega)$ is a source term. The solution can be expanded in the right QNMs as $\psi^{(\mathrm{R})}(\bm{r},\omega) = \sum_n \alpha_n(\omega) \psi^{(\mathrm{R})}_n(\bm{r})$. Inserting this expansion into the previous equation and projecting both sides onto a dual QNM using the inner product in Eq.~\eqref{eqn:QNMs_inner_product}, we obtain:
\begin{align}\label{eqn:alpha_coeff_general}
    \alpha_n(\omega) = -\frac{1}{\omega - \tilde{\omega}_n} \langle \psi^{(\mathrm{L})}_n | S(\omega) \rangle_A \ .
\end{align}
For a general electric current density $\bm{j}(\bm{r},\omega)$ located inside the unmapped domain $\Omega\setminus\Omega_\mathrm{PML}$, so that $S(\omega) = (0, \mathrm{i}\varepsilon^{-1} \bm{j}(\bm{r},\omega), 0, 0)^\mathrm{T}$, this leads to the expression
\begin{align}\label{eqn:alpha_coeff_j}
    \alpha_n(\omega) = -\frac{\mathrm{i}}{\omega - \tilde{\omega}_n}\int_{\Omega_\mathrm{PML}}&\mathrm{d}^3r\, \bm{E}^{(\mathrm{L})}_n(\bm{r}) \cdot \bm{j}(\bm{r},\omega) \ ,
\end{align}
while for a point electric/magnetic dipole emitter at location $\bm{r}_\mathrm{e}$ with $\bm{j}(\bm{r},\omega) = -\mathrm{i}\omega \bm{d}\delta(\bm{r}-\bm{r}_\mathrm{e}) + \nabla \delta(\bm{r}-\bm{r}_\mathrm{e})\times \bm{m}$, we obtain
\begin{align}\label{eqn:dipole_coeff}
    \alpha_n(\omega) = -\frac{\omega}{\omega - \tilde{\omega}_n} \bm{E}^{(\mathrm{L})}_n(\bm{r}_\mathrm{e})\cdot \bm{d}  +\frac{\omega}{\omega - \tilde{\omega}_n}\bm{B}^{(\mathrm{L})}_n(\bm{r}_\mathrm{e})\cdot \bm{m}   \ .
\end{align}
Classically, the spontaneous decay rate $\Gamma = \sum_n \Gamma_n = \sum_n \Gamma_n^\mathrm{e} +  \Gamma_n^\mathrm{m}$ can be calculated from the field generated by the emitter via $\Gamma_n^\mathrm{e}(\omega) = \frac{2}{\hbar} \mathrm{Im}(\alpha_n(\omega) \bm{d}^* \cdot\bm{E}^{(\mathrm{R})}_n(\bm{r}_\mathrm{e}) )$ and $\Gamma_n^\mathrm{m}(\omega) = \frac{2}{\hbar} \mathrm{Im}(\alpha_n(\omega) \bm{m}^* \cdot\bm{B}^{(\mathrm{R})}_n(\bm{r}_\mathrm{e}) )$. Inserting Eq.~\eqref{eqn:dipole_coeff} leads to
\begin{align}\label{eqn:classical_decay_QNMs}
    \Gamma_n &= \frac{2}{\hbar} \mathrm{Re}(\tilde{g}_n^2) \frac{\gamma_n \omega}{(\omega-\omega_n)^2 + \gamma_n^2} \left( 1 - \frac{\mathrm{Im}(\tilde{g}^2)}{\mathrm{Re}(\tilde{g}^2)} \frac{\omega-\omega_n}{\gamma_n} \right) \notag \\
    &\approx \frac{2}{\hbar} \mathrm{Re}(\tilde{g}_n^2) \frac{\gamma_n \omega_n}{(\omega-\omega_n)^2 + \gamma_n^2} \left( 1 - \frac{\mathrm{Im}(\tilde{g}^2)}{\mathrm{Re}(\tilde{g}^2)} \frac{\omega-\omega_n}{\gamma_n} \right) \notag \\
    &= \frac{2}{\hbar} \mathrm{Re}(g_n^2) \frac{\gamma_n}{(\omega-\omega_n)^2 + \gamma_n^2} \left( 1 - \frac{\mathrm{Im}(g^2)}{\mathrm{Re}(g^2)} \frac{\omega-\omega_n}{\gamma_n} \right) \ ,
\end{align}
with $g^2 = \tilde{g}^2\omega_n =  \omega_n(\bm{E}^{(\mathrm{L})}_n(\bm{r}_\mathrm{e})\cdot \bm{d} - \bm{B}^{(\mathrm{L})}_n(\bm{r})\cdot\bm{m})  (\bm{E}^{(\mathrm{R})}_n(\bm{r}_\mathrm{e})\cdot \bm{d}^* + \bm{B}^{(\mathrm{R})}_n(\bm{r})\cdot\bm{m}^*) $. In the second line, we approximated $\omega \approx \omega_n$ in the numerator near the resonance. The first term in Eq.~\eqref{eqn:classical_decay_QNMs} is the usual Lorentzian contribution, similar to the classical result obtained by Purcell; the second term encodes non-Lorentzian lineshapes stemming from overlapping and interfering QNMs as well as from non-Markovian effects. In the single-mode case, $g$ can be used as the coupling constant of a quantum emitter with transition dipole moment $\bm{d}/\bm{m}$ in the interaction Hamiltonian $\hat{H}_\mathrm{int} = g \hat{a}\sigma^+ + \mathrm{h.c.}$.

From Eq.~\eqref{eqn:dipole_coeff}, we also obtain an expansion for the dyadic Green's function of the electric field, defined by $\nabla\times\mu^{-1}\nabla\times G(\bm{r},\bm{r}',\omega) - (\omega^2/c^2) \varepsilon G(\bm{r},\bm{r}',\omega) = \delta(\bm{r}-\bm{r}')$:
\begin{align}\label{eqn:Greens_func_expansion}
    G(\bm{r},\bm{r}',\omega) = \varepsilon_0 c^2\sum_n \frac{1}{\omega(\omega-\tilde{\omega}_n)} \bm{E}^{(\mathrm{R})}_n(\bm{r}) \otimes\bm{E}^{(\mathrm{L})}_n(\bm{r}') \ ,
\end{align}
from which Purcell factors and other significant quantities can be obtained. Another important case is the excitation of the modes by an incoming wave field $(\bm{E}_\mathrm{inc}(\bm{r},\omega), \bm{H}_\mathrm{inc}(\bm{r},\omega))^\mathrm{T}$, satisfying Maxwell's equations
\begin{align}\label{classicalQNM}
    \begin{pmatrix}
    0 & -\mathrm{i}{\mu_\mathrm{b}}^{-1}\nabla\times \\
    \mathrm{i}{\varepsilon_\mathrm{b}}^{-1}\nabla\times & 0 \\
    \end{pmatrix}
    \begin{pmatrix}
    \bm{H}_\mathrm{inc} \\
    \bm{E}_\mathrm{inc}
    \end{pmatrix} 
    = \omega
    \begin{pmatrix}
    \bm{H}_\mathrm{inc} \\
    \bm{E}_\mathrm{inc}
    \end{pmatrix}
\end{align}
in the background medium $(\varepsilon_\mathrm{b},\mu_\mathrm{b})$. After separating the scattered field $(\bm{E}^{(\mathrm{R})}, \bm{H}^{(\mathrm{R})}) = (\bm{E}_\mathrm{inc}, \bm{H}_\mathrm{inc}) + (\bm{E}^{(\mathrm{R})}_\mathrm{sca}, \bm{H}^{(\mathrm{R})}_\mathrm{sca})$, we can expand $(\bm{E}^{(\mathrm{R})}_\mathrm{sca}, \bm{H}^{(\mathrm{R})}_\mathrm{sca})$ in the right QNMs. By using Eq.~\eqref{eqn:alpha_coeff_j} with $\bm{j} = -\mathrm{i}\omega (\varepsilon - \varepsilon_\mathrm{b}) \bm{E}_\mathrm{inc} - \frac{1}{\mathrm{i}\omega}\nabla \times (\mu^{-1}(\omega) - \mu_\mathrm{b}^{-1}) \nabla\times \bm{E}_\mathrm{inc}$, we can ascertain the formula for the excitation coefficients of the scattered field:
\begin{align}
    \alpha_n(\omega) = -\frac{1}{\omega(\omega - \tilde{\omega}_n)} \int_{\Omega_\mathrm{res}}\mathrm{d}^3 r\, \bigg[ \omega^2\bm{E}^{(\mathrm{L})}_n\left(\varepsilon-\varepsilon_\mathrm{b}\right)\bm{E}_\mathrm{inc}  \notag \\
    - (\nabla\times\bm{E}^{(\mathrm{L})}_n) \left(\mu^{-1}(\omega)-\mu_\mathrm{b}^{-1} \right)(\nabla\times\bm{E}_\mathrm{inc}) \bigg]
\end{align}
In the presence of the exchange interaction, the second term cannot be reduced to the action of simple derivative operators as in Eq.~\eqref{eqn:QNM_norm_1}. Its evaluation is therefore non-trivial, and requires the spatial convolution in Eq.~\eqref{eqn:mu_spatial_convolution} with an analytic form or the constructed spatial representation of $\mu(\bm{r},\bm{r}',\omega)$ in Eq.~\eqref{eqn:mu_spatial_expansion}. How good that approximation is and how the presence of exchange affects the modal spectrum are important questions to investigate, but lie beyond the scope of the present investigation. Nevertheless, the QNMs themselves, their normalization, the Green's function and the system's response to local sources can still be ascertained without constructing $\mu(\bm{r},\bm{r}',\omega)$.

It is important to note that the normalization of the left/right QNMs $\tilde{\bm{E}}_n^{(\mathrm{L/R})}$ is not unique. Indeed, a constant rescaling $\tilde{\bm{E}}_n^{(\mathrm{L})} \rightarrow a\tilde{\bm{E}}_n^{(\mathrm{L})}$ in the normalization integral of Eq.~\eqref{eqn:QNM_norm_1} would transform the normalized left/right QNM in the manner $\tilde{\bm{E}}_n^{(\mathrm{L})} \rightarrow \sqrt{a} \tilde{\bm{E}}_n^{(\mathrm{L})}$, $\tilde{\bm{E}}_n^{(\mathrm{R})} \rightarrow \frac{1}{\sqrt{a}} \tilde{\bm{E}}_n^{(\mathrm{R})}$. However, all of the physically significant predictions, like the expanded scattered field, the Green's function and the coupling constants of QEs, always involve products of $\tilde{\bm{E}}_n^{(\mathrm{L/R})}$, rendering them invariant under the aforementioned rescaling.

\section{\label{Sec:Quantization_QNMs} QNM Quantization for Non-reciprocal Magnetic Resonators at Finite Temperatures}

In this section, we briefly review the theory of MQED in non-reciprocal media, before introducing a quantization scheme (in the spirit of Refs.~\cite{franke2019quantization,meschede2026quantum}) for the regularized QNMs of arbitrary, 3D, linear, dissipative (with possible gain contributions) and dispersive, magnetic resonators. Furthermore, we extend our zero-temperature quantum theory to finite temperatures, which constitutes a particularly relevant generalization for cavity quantum magnonics given its characteristic RF and MW spectral domains.

\subsection{Macroscopic QED in Non-Reciprocal Media}

We now generalize the PML-based QNM quantization scheme from Ref.~\cite{meschede2026quantum} to non-reciprocal magnetic resonators, which are describable within the classical framework of \S~\ref{Sec:Classic_QNMs}. The quantization is based on MQED\cite{scheel1998qed,scheel2008macroscopic} and its generalization to non-local and non-reciprocal media\cite{buhmann2012macroscopic}, treating the quantization of the electromagnetic field in the presence of spatially inhomogeneous, lossy/amplifying, dispersive media utilizing the Green's function formalism. For a general non-local conductivity tensor $\bm{Q}(\bm{r},\bm{r}',\omega)$ the frequency components of the electric field operator $\hat{\bm{E}}(\bm{r},\omega)$ satisfy the equation
\begin{align}\label{eqn:MQED_E_field_eq}
    \left[\nabla \times\nabla \times - \frac{\omega^2}{c^2}\right] \hat{\bm{E}}(\bm{r},\omega) - \mathrm{i}\mu_0\omega [Q(\omega)\star \hat{\bm{E}}(\omega)](\bm{r}) \notag \\
    = \mathrm{i}\mu_0\omega \hat{\bm{j}}_\mathrm{\mathrm{F}}(\bm{r},\omega) \ , 
\end{align}
with a fluctuation or noise current density operator
\begin{align}\label{eqn:j_F_general}
    \hat{\bm{j}}_\mathrm{\mathrm{F}}(\bm{r},\omega) = \sqrt{\frac{\hbar\omega}{\pi}} [\mathrm{Re}(Q(\omega))^\frac{1}{2}\star \hat{\bm{f}}(\omega)](\bm{r}) \ , 
\end{align}
in accordance with the fluctuation-dissipation theorem. For simplicity, we omit the superscript $(\mathrm{R})$ on the quantum mechanical operators. The operators $\hat{\bm{f}}^{(\dagger)}(\mathbf{r},\omega)$ are bosonic annihilation and creation operators of the field-dressed matter excitations. These modes are constructed to diagonalize the coupled field-matter Hamiltonian, such that it can be written in the form $\hat{H} = \int\mathrm{d}\omega\int\mathrm{d}^3r\, \hbar\omega \hat{\bm{f}}^{\dagger}(\mathbf{r},\omega) \hat{\bm{f}}(\mathbf{r},\omega)$.
For non-reciprocal and non-local tensors, with $Q(\bm{r},\bm{r}',\omega) \neq Q^\mathrm{T}(\bm{r}',\bm{r},\omega)$, the real and imaginary parts are defined as
$\mathrm{Re}(Q(\bm{r},\bm{r}',\omega)) = ( Q(\bm{r},\bm{r}',\omega) + Q^\dagger(\bm{r}',\bm{r},\omega))/2$ and $\mathrm{Im}(Q(\bm{r},\bm{r}',\omega)) = ( Q(\bm{r},\bm{r}',\omega) - Q^\dagger(\bm{r}',\bm{r},\omega))/2i$. The (anti-)hermitian part of the corresponding integral operator thus satisfies
\begin{align}\label{eqn:Re_Im_tensor_relation}
    \mathrm{Re/Im}(Q(\bm{r}',\bm{r},\omega))_{j i}^* = \mathrm{Re/Im}(Q(\bm{r},\bm{r}',\omega))_{i j} \ ,
\end{align}
and should not be confused with the componentwise real and imaginary parts. In our case, we can express the conductivity in terms of the permittivity and permeability, and decompose the electric/magnetic contributions, allowing us to write the conductivity and fluctuation current as
\begin{align}
    Q(\bm{r},\bm{r}',\omega) = \frac{1}{\mathrm{i}\mu_0\omega} \nabla\times (\tilde{\mu}^{-1}(\bm{r},\bm{r}',\omega) - 1) \times \overset{\leftarrow}{\nabla} \notag \\
    - \mathrm{i}\varepsilon_0\omega\tilde{\varepsilon}(\bm{r},\omega) \delta(\bm{r}-\bm{r}')
\end{align}

The permeability in Eq.~\eqref{eqn:mu_tensor_general} satisfies the Schwarz reflection principle in both $\omega$- and $\bm{k}$-space (\textit{i.e.}, $\mu^*(\omega,\bm{k}) = \mu(-\omega^*,-\bm{k}^*)$), ensuring real-valued response functions in both time and real space. Furthermore, it is analytic in the upper $\omega$ half-plane and admits a finite limit for $|\omega| \rightarrow \infty$, thus satisfying the Kramers-Kronig relations and ensuring causality.

Using the dyadic Green's function of Eq.~\eqref{eqn:MQED_E_field_eq}, the electric field operator in the Schr{\"o}dinger picture can be written as
\begin{align}\label{eqn:E_field_G_expansion_general}
    \hat{\bm{E}}(\bm{r},\omega) &= \mathrm{i}\mu_0 \int_0^\infty\mathrm{d}\omega\, \omega [G(\omega)\star \hat{\bm{j}}_\mathrm{\mathrm{F}}(\omega)](\bm{r}) + \mathrm{h.c.} \notag \\
     &= \hat{\bm{E}}^{(+)}(\bm{r}) + \hat{\bm{E}}^{(-)}(\bm{r}) 
\end{align}
where the decomposition into positive and negative frequency components is facilitated by the Schwarz reflection principle. In Ref.~\cite{buhmann2012macroscopic}, it was shown that an important identity for the imaginary part of the Green's function still holds for non-reciprocal media:
\begin{align}\label{eqn:Im_G_identity}
    \mu_0\omega \int\mathrm{d}^3 s\int\mathrm{d}^3 s'\, G(\bm{r},\bm{s},\omega) \mathrm{Re}(Q(\bm{s},\bm{s}',\omega))G^\dagger(\bm{r}',\bm{s}',\omega) \notag \\
    = \mathrm{Im}(G(\bm{r},\bm{r}',\omega)) \ .
\end{align}
Employing this identity, the field expansion in Eq.~\eqref{eqn:E_field_G_expansion_general} and the fluctuation current density in Eq.~\eqref{eqn:j_F_general}, the vacuum correlation function of the electric field can be ascertained in the form
\begin{align}\label{eqn:corr_func_Im_G}
    \langle  \hat{E}^{(+)}_i(\bm{r}) \hat{E}^{(-)}_j(\bm{r}') \rangle_0  =  \frac{\hbar\mu_0}{\pi} \int_0^\infty \mathrm{d}\omega\,\omega^2 \mathrm{Im}(G(\bm{r},\bm{r}',\omega))_{i j} \ .
\end{align}
The vacuum correlation function has the same form as in the reciprocal case, but crucially, the non-reciprocity manifests in the loss of a symmetry property: $\langle  \hat{E}^{(+)}_i(\bm{r}) \hat{E}^{(-)}_j(\bm{r}') \rangle_0  \neq \langle  \hat{E}^{(+)}_j(\bm{r}') \hat{E}^{(-)}_i(\bm{r}) \rangle_0 $, (\textit{i.e.}, directional quantum correlations without invariance under exchange of positions and directions). The effect of this can be seen, for example, in the interaction of two (or more) dipolar emitters, which is essentially governed by the spectral density matrix $K_{i j} \propto \bm{d}_i^* \mathrm{Im}(G(\bm{r}_i, \bm{r}_j,\omega)) \bm{d}_j$\cite{dung2000mqed}. In the non-reciprocal case, it is possible that $|K_{12}| \neq |K_{21}|$ (\textit{i.e.}, the transfer of excitation between the emitters depends on the direction). For circularly polarized emitters, this implies $|K_{++}| \neq |K_{--}|$, meaning that the interaction of the emitters depends on handedness, since the bias field selects a sense of spin precession in the magnetic medium. 

From the hermiticity of the imaginary part of the Green's function in Eq.~\eqref{eqn:Re_Im_tensor_relation}, it only follows that $\langle  \hat{E}^{(+)}_i(\bm{r}) \hat{E}^{(-)}_j(\bm{r}') \rangle_0  = \langle  \hat{E}^{(+)}_j(\bm{r}') \hat{E}^{(-)}_i(\bm{r}) \rangle_0^* $. 
Using the right-hand side of Eq.~\eqref{eqn:Im_G_identity} and the QNM expansion of the Green's function in Eq.~\eqref{eqn:Greens_func_expansion}, the vacuum correlation function can be expressed as
\begin{align}\label{eqn:corr_func_QNM_expansion}
    \langle  \hat{E}^{(+)}_i(\bm{r}) \hat{E}^{(-)}_j(\bm{r}') \rangle_0  &= \left[ \hat{E
    }^{(+)}_i(\bm{r}), \hat{E}^{(-)}_j(\bm{r}') \right] \notag \\
    &=\sum_{n n'} P_{n n'} \tilde{E}^{(\mathrm{R})}_{n, i}(\bm{r})\tilde{E}^{(\mathrm{R})*}_{n', j}(\bm{r}') \ ,
\end{align}
with
\begin{align}\label{eqn:P_mat}
    P_{n n'} = \frac{\hbar\mu_0}{\pi} \int_0^\infty\mathrm{d}\omega\,\omega^2\int\mathrm{d}^3s\, \frac{\omega^2}{c^2} \tilde{\bm{E}}^{(\mathrm{L})}_n(\bm{s}) \mathrm{Im}(\tilde{\varepsilon}(\bm{s},\omega) )\tilde{\bm{E}}^{(\mathrm{L})*}_{n'}(\bm{s}) \notag \\
    - \nabla\times\tilde{\bm{E}}^{(\mathrm{L})}_n(\bm{s}) \left[\mathrm{Im}(\tilde{\mu}(\omega) )\star  (\nabla\times\tilde{\bm{E}}^{(\mathrm{L})*}_{n'})\right](\bm{s}) \ .
\end{align}
At this point, it is evident that both left and right QNMs are required for the quantization scheme, although the $P$-matrix in Eq.~\eqref{eqn:P_mat}, needed for the mode symmetrization itself, can be computed solely from the left QNMs. Just as for reciprocal media, we use PMLs to truncate the open space to a finite domain $\Omega_\mathrm{PML}$, and the spatial integral in Eq.~\eqref{eqn:P_mat} is then performed over all of $\Omega_\mathrm{PML}$. As discussed in Ref.~\cite{meschede2026quantum}, the PMLs also contain gain components (\textit{i.e.}, components of the effective material parameters $(\tilde{\mu},\tilde{\varepsilon})$ with negative imaginary parts). If the resonator itself is comprised of gain media, the noise current would be modified and the MQED creation/annihilation operators $\hat{\bm{f}}^{(\dagger)}(\bm{r},\omega)$ would interchange roles in the amplifying regions\cite{raabe2008qed,franke2022coupled,meschede2026quantum}. However, the gain components of the PMLs are not explicitly taken into account, but simply treated as an effective reduced loss in the integral of Eq.~\eqref{eqn:P_mat}, so that the QNM expansion of the vacuum correlation function in Eq.~\eqref{eqn:corr_func_QNM_expansion} reproduces the correct MQED result in Eq.~\eqref{eqn:corr_func_Im_G}.

\subsection{QNM Quantization}

Analogous to the reciprocal case, we now introduce a QNM expansion for the electric field operator $\hat{\bm{E}}^{(+)}(\bm{r}) = \sum_n \tilde{\bm{E}}^{(\mathrm{R})}_n(\bm{r}) \hat{\xi}_n$\cite{franke2019quantization,franke2022coupled,meschede2026quantum}. Inserting this expansion into the commutator and comparing it to Eq.~\eqref{eqn:corr_func_QNM_expansion} shows that $\left[ \hat{\xi}_n, \hat{\xi}_{n'}^\dagger\right] = P_{n n'}$ (\textit{i.e.}, non-bosonic commutation relations arising from the non-hermitian nature of the system). The symmetrization transformation $\hat{a}_n = P^{-\frac{1}{2}}_{n n'} \hat{\xi}_{n'}$ yields bosonic modal operators $\hat{a}_n$ satisfying $[\hat{a}_n, \hat{a}_{n'}^\dagger] = \delta_{n n'}$, and the expansion $\hat{\bm{E}}^{(+)}(\bm{r}) = \sum_n \tilde{\bm{E}}^{(\mathrm{R},\mathrm{sym})}_n(\bm{r}) \hat{a}_n$ with the symmetrized QNMs $\tilde{\bm{E}}^{(\mathrm{R},\mathrm{sym})}_n(\bm{r}) = P^{\frac{1}{2}}_{n' n} \tilde{\bm{E}}^{(\mathrm{R})}_{n'}(\bm{r})$. The modal operators themselves can be written in the form
\begin{align}
    \hat{a}_n = \mathrm{i}\mu_0 &P^{-\frac{1}{2}}_{n n'} \int_0^\infty\mathrm{d}\omega\,\omega A_{n'}(\omega) \int\mathrm{d}^3r\, \bigg[ \notag \\ \omega &\sqrt{\frac{\hbar\varepsilon_0}{\pi} \mathrm{Im}(\varepsilon(\bm{r},\omega))} \tilde{\bm{E}}^{(\mathrm{L})}_{n'}(\bm{r})\cdot \hat{\bm{f}}_\mathrm{e}(\bm{r},\omega)  \\
    +&\sqrt{\frac{\hbar}{\mu_0\pi}} [ \mathrm{Im}(\mu^{-1})^\frac{1}{2}\star (\nabla\times \tilde{\bm{E}}^{(\mathrm{L})}_{n'}) ](\bm{r}) \cdot \hat{\bm{f}}_\mathrm{m}(\bm{r},\omega) \bigg] \ . \notag
\end{align}
Namely, as superpositions of the continuum MQED modes constituting the (symmetrized) QNMs. They satisfy the Heisenberg-Langevin equation
\begin{align}
    \frac{\mathrm{d}\hat{a}_n}{\mathrm{d}t} = \frac{\mathrm{i}}{\hbar} \left[ \hat{H}_\mathrm{eff}, \hat{a}_n \right] - \chi^{-}_{n n'}\hat{a}_{n'} + \hat{F}_n(t) \ , 
\end{align}
with the effective Hamiltonian $H_\mathrm{eff} = \hbar \chi^+_{n n'} \hat{a}^\dagger_n \hat{a}_{n'}$, the hermitian/antihermitian parts $\chi_{n n'}^\pm  =\frac{(\mathrm{i})}{2} ( \chi_{n n'} \pm \chi^\dagger_{n n'})$ of the matrix $\chi_{n n'} = P^{-\frac{1}{2}}_{n m} \tilde{\omega}_m P^{+\frac{1}{2}}_{m n'}$, and the Langevin noise term
\begin{align}
    \hat{F}_n(t) = &\mathrm{i}\mu_0 P^{-\frac{1}{2}}_{n n'} \int_0^\infty\mathrm{d}\omega\,\omega (\tilde{\omega}_{n'} - \omega)A_{n'}(\omega) \int\mathrm{d}^3r\, \bigg[ \notag \\ \omega &\sqrt{\frac{\hbar\varepsilon_0}{\pi} \mathrm{Im}(\varepsilon(\bm{r},\omega))} \tilde{\bm{E}}^{(\mathrm{L})}_{n'}(\bm{r})\cdot \hat{\bm{f}}_\mathrm{e}(\bm{r},\omega)  \\
    +&\sqrt{\frac{\hbar}{\mu_0\pi}} [ \mathrm{Im}(\mu^{-1})^\frac{1}{2}\star (\nabla\times \tilde{\bm{E}}^{(\mathrm{L})}_{n'}) ](\bm{r}) \cdot \hat{\bm{f}}_\mathrm{m}(\bm{r},\omega) \bigg] \ . \notag
\end{align}
Approximating this term as white noise and invoking It{\^{o}}-Stratonovich calculus leads to an effective Markovian master equation\cite{gardiner1985input, franke2019quantization, franke2022coupled},
\begin{align}\label{eqn:QNM_master_eq}
    \frac{\mathrm{d}\rho}{\mathrm{d}t} = -\frac{\mathrm{i}}{\hbar} \left[\hat{H}_\mathrm{eff}, \rho \right] + \chi^{-}_{n n'} \mathcal{D}_{n n'} \rho \ ,
\end{align}
with the dissipators $\mathcal{D}_{n n'}\rho = 2 \hat{a}_{n'}\rho \hat{a}_n^\dagger - \{\rho, \hat{a}_n^\dagger \hat{a}_{n'}\}$. The symmetrized QNMs couple both coherently and dissipatively via the effective Hamiltonian and the dissipators, respectively, which facilitates a rigorous treatment of dissipative photon-magnon coupling.

As in the classical case, we can show that the physical predictions of the theory are independent of any rescaling of the left and right QNMs. Rescaling of the QNMs before their normalization as $\tilde{\bm{E}}^{(\mathrm{L/R})}(\bm{r}) \rightarrow a_{\mathrm{L/R}} \tilde{\bm{E}}^{(\mathrm{L/R})}(\bm{r})$ leads to the normalized QNMs $\tilde{\bm{E}}^{(\mathrm{L/R})}(\bm{r}) \rightarrow U^{(\mathrm{L/R})}\tilde{\bm{E}}^{(\mathrm{L/R})}(\bm{r})$, with transformation matrices $U_{n n'}^{(\mathrm{L/R})}=a^{(\mathrm{L/R})}_n/(a_n^{(\mathrm{L})}a_n^{(\mathrm{R})})^\frac{1}{2} \delta_{n n'}$, being non-unitary by themselves, but together satisfying $U^{(\mathrm{L})}U^{(\mathrm{R})} = U^{(\mathrm{R})\dagger}U^{(\mathrm{L})\dagger} = \bm{I}$. Correspondingly, the $P$-matrix transforms as $P \rightarrow U^{(\mathrm{L})} P U^{(\mathrm{L})\dagger}$. It can be evidenced that under these transformations, the vacuum correlation function remains invariant:
\begin{align}
    \langle  E^{(+)}_i(\bm{r}) E^{(-)}_j(\bm{r}') \rangle_0 &= \tilde{\bm{E}}^{(\mathrm{R})}(\bm{r})P \tilde{\bm{E}}^{(\mathrm{R})*}(\bm{r}') \notag \\
    \rightarrow &\tilde{\bm{E}}^{(\mathrm{R})}(\bm{r})U^{(\mathrm{R})}U^{(\mathrm{L})} P U^{(\mathrm{L})\dagger} U^{(\mathrm{R})\dagger}\tilde{\bm{E}}^{(\mathrm{R})*} \notag \\
    &= \langle  E^{(+)}_i(\bm{r}) E^{(-)}_j(\bm{r}') \rangle_0 \ . 
\end{align}
The symmetrized QNMs transform as
\begin{align}
    \tilde{\bm{E}}^{(\mathrm{R},\mathrm{sym})} \rightarrow (U^{(\mathrm{L})\dagger} P^\mathrm{T} U^{(\mathrm{L})} )^\frac{1}{2} U^{(\mathrm{R})}  \tilde{\bm{E}}^{(\mathrm{R})} = U^{\mathrm{sym}} \tilde{\bm{E}}^{(\mathrm{R},\mathrm{sym})} \ ,
\end{align}
with $U^{\mathrm{sym}} = (U^{(\mathrm{L})\dagger} P^\mathrm{T} U^{(\mathrm{L})} )^\frac{1}{2} U^{(\mathrm{R})}(P^\mathrm{T})^{-\frac{1}{2}}$ being a unitary transformation:
\begin{align}
    U^{\mathrm{sym}\dagger} U^{\mathrm{sym}} &= (P^\mathrm{T})^{-\frac{1}{2}}U^{(\mathrm{R})\dagger}U^{(\mathrm{L})\dagger} P^\mathrm{T}  U^{(\mathrm{L})}U^{(\mathrm{R})}    (P^\mathrm{T})^{-\frac{1}{2}} \notag \\
    &= (P^\mathrm{T})^{-\frac{1}{2}} P^\mathrm{T}(P^\mathrm{T})^{-\frac{1}{2}} = \bm{I} \ .
\end{align}
Although the scaling generally leads to a non-unitary transformation of the QNMs, the symmetrized QNMs always transform unitarily (\textit{i.e.}, they are uniquely determined up to a unitary transformation). The bosonic modal operators transform as $\hat{a}_n \rightarrow U^{\mathrm{sym}*} \hat{a}_n$, and from its definition, we can show that $\chi$ transforms accordingly as $\chi \rightarrow U^{\mathrm{sym}*} \chi (U^{\mathrm{sym}*})^{-1}$. In conjunction with the transformation of the noise operator, we find that the QNM master equation~\eqref{eqn:QNM_master_eq} and thus the quantum dynamics remain invariant under these transformations, as demanded.

\subsection{Finite Temperature}
\label{subsec:finite_temp}
So far, QNM quantization has been considered only at zero temperature\cite{franke2019quantization,meschede2026quantum}. This has proven adequate for the photonic and plasmonic structures to which the formalism has most commonly been applied, where the bosonic mode frequencies lie in the optical range and thermal occupation is negligible.
In quantum nano-optical scenarios, finite-temperature effects are often accounted for simply by an increased pure dephasing rate of the QEs embedded in the system. In contrast, since the excitation frequencies of magnonic modes and spin/SC qubits are often in the MHz to GHz range, thermal excitation of the modes is significantly more pronounced, even at cryogenic temperatures, and must be taken into account.

In the general theory of open quantum systems, thermal effects are typically described by assuming that the mode is coupled to a continuous bath in thermal equilibrium, leading to an incoherent pumping term in the corresponding master equation of the form $\mathcal{D}_\mathrm{p}\rho = \gamma n_T ( 2 \hat{a}^\dagger\rho \hat{a} - \{\rho, \hat{a} \hat{a}^\dagger \})$, with the Bose-Einstein distribution $n_T = 1/(\exp(\hbar\omega/k T) - 1)$, while the incoherent loss term becomes $\mathcal{D}_\mathrm{l}\rho = \gamma (1 + n_T) ( 2 \hat{a}\rho \hat{a}^\dagger - \{\rho, \hat{a}^\dagger \hat{a}\})$, including both spontaneous and thermally stimulated emission. For the expectation value of the bosonic number operator, this leads to the equation ${\mathrm{d}} \langle\hat{n}(t)\rangle/dt = -2\gamma( \langle\hat{n}(t)\rangle - n_T )$, with solution $\langle\hat{n}(t)\rangle = \langle\hat{n}(0)\rangle \mathrm{e}^{-2\gamma t} + n_T (1 - \mathrm{e}^{-2\gamma t})$, so that the mode relaxes towards the thermal equilibrium boson number $n_T$ with the spontaneous decay rate $\gamma$.  

Building on the QNM quantization discussed earlier in this section, we now derive an effective thermal QNM master equation, in which interference effects arising from the shared bath are rigorously taken into account, spatially dependent temperature profiles $T(\bm{r})$ can be described, and all relevant quantities can be derived and calculated from traditional MQED. Assuming that the MQED continuum modes are in thermal equilibrium, namely $\langle \hat{f}_{\lambda,i}(\bm{r},\omega)\hat{f}^\dagger_{\lambda',j}(\bm{r}',\omega') \rangle = (n_{T(\bm{r})}(\omega)+1) \delta(\omega-\omega')\delta(\bm{r}-\bm{r}') \delta_{i j} \delta_{\lambda \lambda'}
$, the electric field thermal correlation function becomes
\begin{align}\label{eqn:E_field_thermal_corr}
    &\langle  \hat{E}^{(+)}_i(\bm{r}) \hat{E}^{(-)}_j(\bm{r}') \rangle_T  \notag \\
    &=\frac{\hbar \mu_0^2}{\pi}\int_0^\infty \mathrm{d}\omega\, (n_T(\omega)+1)\omega \int\mathrm{d}^3 s\int\mathrm{d}^3 s'\bigg[ \notag \\
    &\hspace{2cm} G(\bm{r},\bm{s},\omega) \mathrm{Re}(Q(\bm{s},\bm{s}',\omega))G^\dagger(\bm{r}',\bm{s}',\omega)\bigg] \notag \\
    &\overset{T(\bm{r})=T}{=}  \frac{\hbar\mu_0}{\pi} \int_0^\infty \mathrm{d}\omega\,  (n_T(\omega)+1)\omega^2 \mathrm{Im}(G(\bm{r},\bm{r}',\omega))_{i j} \ , \notag \\
    &= ( P_{n n'} + P_{T, n n'} )\tilde{\bm{E}}_{n,i}(\bm{r})\tilde{\bm{E}}^*_{n,j}(\bm{r}') \ ,
\end{align}
where the penultimate equality holds under the assumption of a spatially uniform temperature distribution; note that the common representation through the imaginary part of the Green's function can no longer be used in the case of a spatially varying $T(\bm{r})$. For the last equality, we have simply used Eq.~\eqref{eqn:Greens_func_expansion} and the (hermitian) thermal $P$-matrix $P^T_{n n'}$ is defined as in Eq.~\eqref{eqn:P_mat}, with the additional Bose-Einstein factor $n_T(\omega)$ in the frequency integral. Using the expansion of the electric field operator in terms of the symmetrized QNMs, this expectation value can also be expressed as
\begin{align}
    \langle  \hat{E}^{(+)}_i(\bm{r}) \hat{E}^{(-)}_j(&\bm{r}') \rangle_\mathrm{T} = \tilde{E}^\mathrm{(R,sym)}_{n,i}(\bm{r})\tilde{E}^\mathrm{(R,sym)*}_{n',j}(\bm{r}')\langle  \hat{a}_{n}\hat{a}_{n'}^\dagger \rangle_T \notag \\
    &= P^\frac{1}{2}_{m n} P^\frac{1}{2}_{n' m'} \tilde{E}^\mathrm{(R)}_{m,i}(\bm{r}) \tilde{E}^\mathrm{(R)*}_{m',j}(\bm{r}') \langle  \hat{a}_{n}\hat{a}_{n'}^\dagger \rangle_T \notag \\
    &\overset{!}{=} (P_{m m'} + P_{T, m m'} )\tilde{E}_{m,i}(\bm{r})\tilde{E}^*_{m',j}(\bm{r}') \ .
\end{align}
The modal thermal expectation values are thus given by $\langle \hat{a}_{n}^\dagger \hat{a}_{n'} \rangle_T = P^{-\frac{1}{2}*}_{n m} P^*_{T,m m'} P^{-\frac{1}{2}*}_{m' n'}$. To derive the thermal QNM master equation, we now make the ansatz
\begin{align}\label{eqn:thermal_master_eq}
    \frac{\mathrm{d}\rho}{\mathrm{d}t} = -\frac{\mathrm{i}}{\hbar} \left[\hat{H}_\mathrm{eff}, \rho \right] + (\chi^{-}_{n n'}  +\chi^{\mathrm{G}}_{n n'}(T) )\mathcal{D}^\mathrm{L}_{n n'} \rho \notag \\
    +\chi^{\mathrm{G}}_{n n'}(T) \mathcal{D}^\mathrm{G}_{n n'} \rho  \ ,
\end{align}
with a modified loss term and an incoherent pumping dissipator $\mathcal{D}^\mathrm{G}_{n n'} \rho = 2\hat{a}_{n}^\dagger \rho \hat{a}_{n'} - \{\rho, \hat{a}_{n'} \hat{a}^\dagger_n \}$ that includes crossing terms arising from the shared bath. From Eq.~\eqref{eqn:thermal_master_eq}, we can derive equations of motion for the expectation values $\langle \hat{a}_n^\dagger \hat{a}_{n'} \rangle$:
\begin{align}
    \frac{\mathrm{d}}{\mathrm{d}t}\langle \hat{a}_n^\dagger \hat{a}_{n'} \rangle = &- (\mathrm{i} \chi^+_{n' m'} + \chi^-_{n' m'}) \langle \hat{a}_n^\dagger \hat{a}_{m'} \rangle \notag \\
    &+(\mathrm{i} \chi^+_{m n} - \chi^-_{m n} )\langle \hat{a}_m^\dagger \hat{a}_{n'} \rangle + 2\chi_{n' n}^\mathrm{G}(T) \ ,
\end{align}
for which the thermal steady-state condition $\frac{\mathrm{d}}{\mathrm{d}t}\langle \hat{a}_n^\dagger \hat{a}_{n'} \rangle \overset{!}{=} 0$ leads to
\begin{align}
    \chi^\mathrm{G}_{n n'}(T) = \frac{\mathrm{i}}{2} \left( \chi_{n m'} \langle \hat{a}^\dagger_{n' } \hat{a}_{m'}\rangle_T - \chi^\dagger_{m n'} \langle \hat{a}^\dagger_{m } \hat{a}_{n}\rangle_T  \right) \ .
\end{align}
Choosing the incoherent gain matrix $\chi^\mathrm{G}_{n n'}(T)$ such that it reproduces the modal thermal steady states $\langle \hat{a}^\dagger_{n} \hat{a}_{n'}\rangle_T$ calculated from MQED leads to
\begin{align}
    \chi^\mathrm{G}(T) &= \frac{\mathrm{i}}{2} \left( \chi P^{-\frac{1}{2}} P_T^*  P^{-\frac{1}{2}} - P^{-\frac{1}{2}} P_T^*  P^{-\frac{1}{2}} \chi^\dagger \right) \notag \\
    &= - \mathrm{Im}( \chi P^{-\frac{1}{2}} P_T  P^{-\frac{1}{2}} ) = - \mathrm{Im}( P^{-\frac{1}{2}} P_T \tilde{\bm{\omega}} P^{-\frac{1}{2}} ) \ ,
\end{align}
with $\tilde{\bm{\omega}} = \mathrm{diag}(\tilde{\omega}_n)$ and $\mathrm{Im}(A) = 1/(2\mathrm{i})(A-A^\dagger)$ being the anti-hermitian part of the matrix.

\section{\label{Sec:Numerical Examples} Numerical Examples}

Complementary to the general formalism introduced in the preceding sections, we now illustrate the predictive capabilities of our quantum QNM methodology by studying, in a spatiotemporally resolved manner, both photon-magnon and magnon-qubit interactions. In \S~\ref{Subsec:1D_examples}, we begin by considering a 1D, half-open magnonic slab, capturing both the fundamental FMR and higher-order modes, as well as their complex interplay. Second, we consider a similar magnonic slab embedded in a photonic cavity, illustrating the ability of our method to predict both coherent and dissipative magnon-photon coupling regimes and their interpolation, together with a fully spatial and temporal characterization of the quantum field dynamics. We extend our approach to a 3D case in \S~\ref{Subsec:3D_examples}, exploring a spherically symmetric magnon-photon hybrid resonator and its interaction with SQs and FQs.

For the purpose of our numerical demonstrations, we adopt the canonical magnonic material YIG\cite{cherepanov1993yig,serga2010yig}, with Gilbert damping parameter $\alpha = 3\cdot 10^{-4}$, saturation magnetization $\mu_0 M_\mathrm{s} = 175\,\mathrm{mT}$ and relative permittivity $\varepsilon_\mathrm{YIG} = 15$. The gyromagnetic ratio is chosen for the spin-only moment of $\mathrm{Fe}^{3+}$, so that $\gamma/(2\pi) = 28\, \mathrm{GHz/T}$. Throughout this section, we shall neglect magnetocrystalline anisotropy and the exchange interaction; whilst the former is negligible in the case of YIG magnonics, the influence of the latter could become significant for nanoscale systems and shall be studied elsewhere.

\subsection{\label{Subsec:1D_examples} 1D Systems}

\paragraph{Magnonic Slab Resonator}

We consider a YIG slab of length $L_\mathrm{YIG}$ in the $z$-direction and of infinite extent in the $x$- and $y$-directions, subject to an out-of-plane bias field $\bm{H}_0  = H_0 \bm{e}_z$ with $H_{0} \hat{=} H_0 - N_z M_\mathrm{s}$, $N_z = 1$. We impose a perfect electrical conductor boundary condition at $z = 0$, such that the slab radiates into a half-open space. The permeability in Eq.~\eqref{eqn:mu_tensor_general} has $\mu_{xx} = \mu_{yy}$ and eigenvalues $\{\mu_{x x} \pm\kappa, 1\}$, with corresponding eigenvectors $\{(1,\pm\mathrm{i},0)^\mathrm{T},(0,0,1)^\mathrm{T} \}$. We consider waves propagating along the $z$-direction and treat the system as 1D. As apparent from Eq.~\eqref{eqn:P_mat}, the left and right circularly polarized eigenmodes do not couple and can be treated separately in the quantization; here, we retain only the left circularly polarized modes, since the FMR occurs for this polarization. The right QNMs have the form $\tilde{\bm{E}}^{(\mathrm{R})}_n(z) = E^{(\mathrm{R})}_n(z) ( 1, \mathrm{i},0)^\mathrm{T}/\sqrt{2}$ and can be calculated from the 1D wave equation $\partial_z (\mu^{-1}_\mathrm{eff}\partial_z E^{(\mathrm{R})}_n) + (\omega^2/c^2) \varepsilon E^{(\mathrm{R})}_n = 0$, with the effective relative permeability $\mu_\mathrm{eff}(z,\omega)=\mu_{xx}(\omega) + \kappa(\omega)$ ($z < L_\mathrm{YIG}$) and $\mu_\mathrm{eff}(z,\omega)=1$ ($z \geq L_\mathrm{YIG}$), together with the imposed boundary condition $E^{(\mathrm{R})}_n(z=0) = 0$. The corresponding left QNMs are given by $\tilde{\bm{E}}^{(\mathrm{L})}_n(z) = E^{(\mathrm{R})}_n(z) ( 1, -\mathrm{i},0)^\mathrm{T}/\sqrt{2}$ (for transposed permeability), and the $P$-matrix elements in Eq.~\eqref{eqn:P_mat} reduce to
\begin{align}
    P_{n n'} = \frac{\hbar \mu_0}{\pi A}  &\int_0^\infty \mathrm{d}\omega\,\omega^2 A_n(\omega) A^*_{n'}(\omega) \int_{\Omega_\mathrm{PML}}\mathrm{d}s\, \bigg[ \notag \\
    &+\frac{\omega^2}{c^2}\, \mathrm{Im}(\tilde{\varepsilon}(s,\omega) ) E^{(\mathrm{R})}_{n}(s) E^{(\mathrm{R}) *}_{n'}(s) \notag \\
    &- \mathrm{Im}(\tilde{\mu}^{-1}_\mathrm{eff}(s,\omega)) \partial_s E^{(\mathrm{R})}_{n}(s) \partial_s E^{(\mathrm{R})*}_{n'}(s) \bigg] \ , 
\end{align}
with the normalization area $A$ in the $xy$-plane. Considering a two-level magnetic QE with frequency $\omega_\mathrm{QE}$ coupled to the resonator, the QE decay rate can be obtained from the QNM master equation in the bad-cavity limit\cite{cirac1992interaction, franke2019quantization, meschede2026quantum}:
\begin{align}
    \Gamma_\mathrm{QE} = \frac{1}{\hbar^2} \mathrm{Re}\left( g_{n} P^{+\frac{1}{2}}_{n n'} \frac{1}{\mathrm{i}(\omega_\mathrm{QE} - \tilde{\omega}^*_{n'} )} P^{-\frac{1}{2}}_{n' n''}  g^*_{n''}   \right) \ .
\end{align}
The QE decay rate can also be calculated exactly from the 1D Green's function of the system via $\Gamma_\mathrm{QE} = (\mu_{0}|m_\mathrm{QE}|^{2}/\hbar A) \lim\limits_{z' \rightarrow z} \partial_z \partial_{z'}\mathrm{Im}(G(z,z',\omega))$, which is known analytically for this system.~Normalization with respect to the 1D free space decay rate $\Gamma^0_\mathrm{QE} = \mu_0 \omega_\mathrm{QE} |m_\mathrm{QE}|^2/(2 \hbar c A)$ yields an $m_\mathrm{QE}$- and $A$-independent magnetic Purcell factor. 

\begin{figure}
    \centering
    \includegraphics[width=1.0\linewidth]{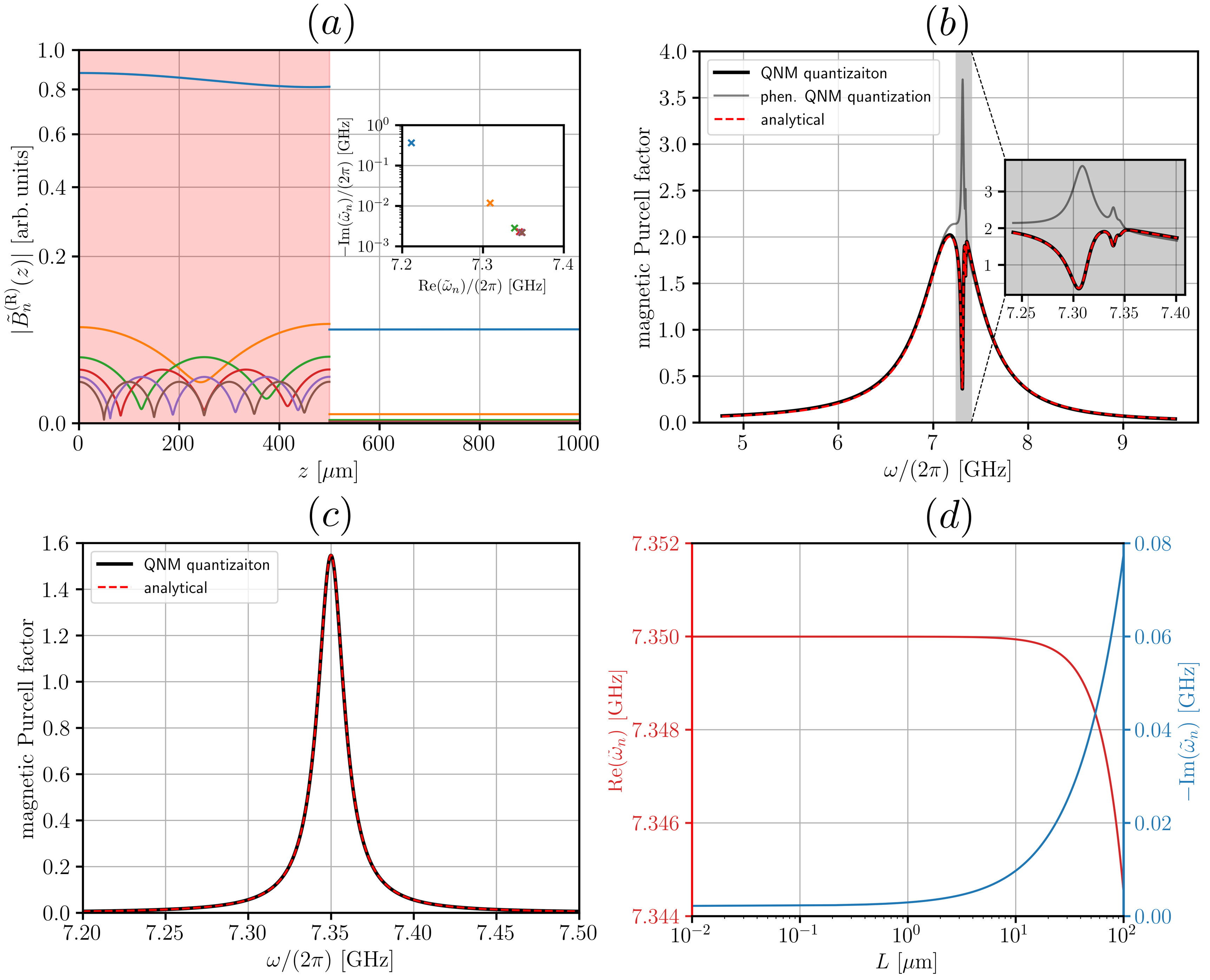}
    \caption{Numerical investigation of a 1D YIG slab in 1D half-open space, magnetized in the positive $z$-direction with $H_0 = 1.5\,\mathrm{M}_s$. (a) Magnetic flux density profiles $|\tilde{B}^{(\mathrm{R})}_n(z)|$ and complex frequencies $\tilde{\omega}_n$ of the lowest left circularly polarized QNMs for a slab of length $L_\mathrm{YIG} = 500\,\mu \mathrm{m}$. (b) Frequency-dependent magnetic Purcell factor for a QE positioned at the open end of the slab ($z=L_\mathrm{YIG}$), obtained from the full quantum QNM theory, the phenomenological QNM quantization scheme and the analytical treatment. (c) Corresponding magnetic Purcell factor for a slab of length $L_\mathrm{YIG} = 10\,\mu \mathrm{m}$, calculated from a single-mode (Kittel) expansion in the quantum QNM theory, compared with the analytical result. (d) Real and imaginary parts of the complex QNM frequency
    of the Kittel mode as a function of $L_\mathrm{YIG}$.}
    \label{fig:1D_slab}
\end{figure}

Fig.~\ref{fig:1D_slab}(a) presents the six lowest complex QNM frequencies $\tilde{\omega}_n$, along with the corresponding magnetic flux density profiles $|B_n(z)|$, for a YIG slab of length $L_\mathrm{YIG}=500\,\mu\mathrm{m}$ magnetized along the positive $z$-direction and subject to a bias field of $H_0 = 1.5\,M_\mathrm{s}$. At the pole of the effective permeability, $\tilde{\omega}_\mathrm{pole} = (\omega_\mathrm{H} - \omega_\mathrm{M})/(1+\mathrm{i}\alpha)$, there is an accumulation point of QNMs (\textit{i.e.}, an infinite set of QNMs with $\tilde{\omega}_n \rightarrow \tilde{\omega}_\mathrm{pole}$), as also appear in plasmonic systems\cite{sauvan2022norm}. The lowest mode is the nearly uniform Kittel mode, which appears to have the lowest $Q$-factor, whereas the higher-order modes are spatially oscillatory. The accumulation point arises from the divergence of the refractive index, so that the YIG slab of fixed length can accommodate an arbitrarily large number of oscillations as the wavelength in the material tends to zero at $\tilde{\omega}_\mathrm{pole}$. In Fig.~\ref{fig:1D_slab}(b), the corresponding Purcell factor for a magnetic QE positioned at the open end of the slab resonator ($z = L_\mathrm{YIG}$) is shown, calculated from both the quantum QNM methodology, retaining the six modes of Fig.~\ref{fig:1D_slab}(a), and the analytical Green's function. The Purcell factors are found to be in excellent quantitative agreement. Notably, the higher-order modes lead to dips rather than additional peaks -- stemming from interference effects in the dissipative coupling of these modes -- which the quantum QNM theory reproduces very well. Our calculated Purcell factors imply that the dynamics of the Kittel mode, and of QEs interacting with it, are directly impacted by the presence of the higher-order modes. In addition, we plot the Purcell factor calculated from the phenomenological quantization of the QNMs, using the QE-QNM coupling constants from Eq.~\eqref{eqn:classical_decay_QNMs} and uncoupled QNMs (\textit{i.e.}, $\hat{H}_\mathrm{eff} = \sum_n \hbar \omega_n \hat{a}^\dagger_n \hat{a}_n$). Evidently, such a phenomenological approach is not capable of reproducing the dips, illustrating the need for a rigorous QNM quantization to capture the interplay between the Kittel and the higher-order modes.

In Fig.~\ref{fig:1D_slab}(c), we present the magnetic Purcell factor for a YIG slab of length $L_\mathrm{YIG}=50\,\mu\mathrm{m}$, comparing the analytical result with a single-mode (Kittel) expansion in the quantum QNM treatment. For this thinner slab, a single-mode expansion already yields very good agreement, indicating that the influence of the higher-order modes becomes negligible for decreasing size of the YIG slab; moreover, although not shown in this plot, we find that the phenomenological quantization protocol suffices to reproduce the analytical result as well. To provide additional insight into the role of the cavity length, we also present the $L_\mathrm{YIG}$-dependence of $\mathrm{Re}(\tilde{\omega}_n)$ and $\mathrm{Im}(\tilde{\omega}_n)$ for the Kittel mode in Fig.~\ref{fig:1D_slab}(d). We observe that $\tilde{\omega}_n$ approaches $\tilde{\omega}_\mathrm{pole}$ as $L_\mathrm{YIG} \rightarrow 0$, and since the effective wavenumber tends to a finite value in this limit, the Kittel mode profile becomes homogeneous inside the YIG slab.
It is worth recalling that we have neglected the exchange interaction, which must be taken into account if the size of the magnetic structure approaches the exchange length, typically within the nanometer range\cite{abo2013definition,xue2024high}, especially for the higher-order modes with non-uniform magnetization profile. The present simple system would provide a useful platform to systematically study the effects of the exchange interaction on the QNMs and their coupling with magnetic QEs at small slab dimensions $L_\mathrm{YIG}$. However, such an investigation shall be reported elsewhere.

\paragraph{Hybrid Magnonic-photonic Slab Resonator}

\begin{figure}
    \centering
    \includegraphics[width=1.0\linewidth]{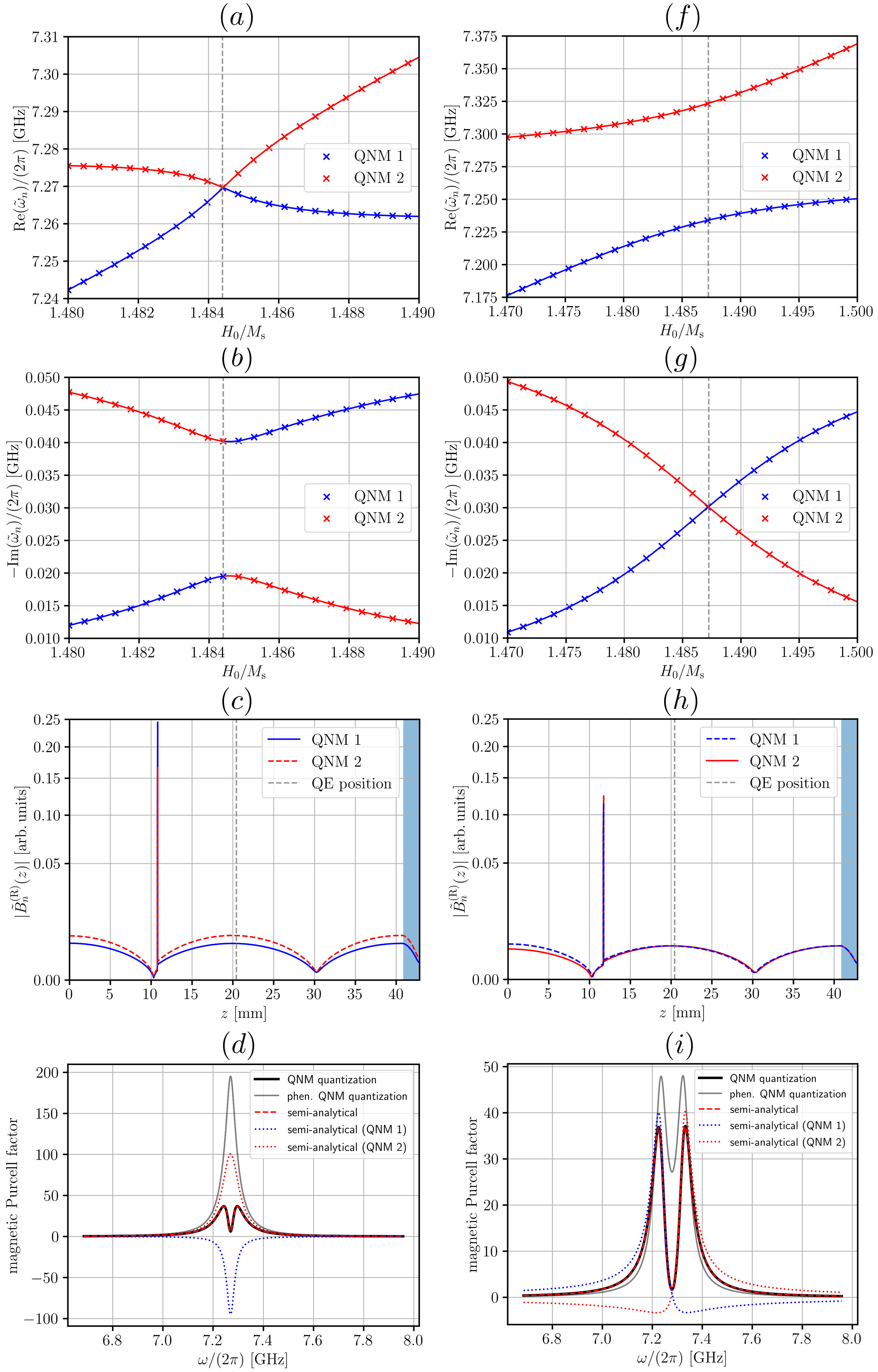}
    \caption{Numerical investigation of a 1D YIG slab of length $L_\mathrm{YIG} = 50\,\mu\mathrm{m}$ embedded in a half-open cavity of length $L_\mathrm{cav} = 40.860\,\mathrm{m m}$, magnetized in the positive $z$-direction. The permittivity of the cavity material is defined by Eq.~\eqref{eqn:1D_permittivity_cavity}. The second-order mode of the cavity is tuned to the Kittel mode of the YIG slab, and those two QNMs are retained. Left column: $L_{\mathrm{air}}=10.753\,\mathrm{mm}$. Right column: $L_{\mathrm{air}}=11.674\,\mathrm{mm}$. (a),(f) Real parts of the QNM frequencies, $\mathrm{Re}(\tilde{\omega}_n)$, as a function of bias field. (b),(g) Imaginary parts of the QNM frequencies, $\mathrm{Im}(\tilde{\omega}_n)$, as a function of bias field. (c),(h) QNM magnetic flux density profiles $|\tilde{B}^{(\mathrm{R})}_n(z)|$ at the bias fields $H_0$ marked by the dashed lines in the panels above. (d),(i) Frequency-dependent magnetic Purcell factor at the QE position marked in (c),(h), obtained from the full quantum QNM theory, the phenomenological QNM quantization scheme and the analytical treatment.}
    \label{fig:1D_slab_in_cavity}
\end{figure}

An even more enriching and stringent case for the application of our quantum QNM theory concerns the problem of magnon-photon coupling in a hybrid configuration, where the YIG slab is embedded in a half-open photonic cavity comprising a high-index dielectric slab ($\varepsilon = 20$). The permittivity distribution is defined by
\begin{align}\label{eqn:1D_permittivity_cavity}
    \varepsilon(z,\omega) = \begin{cases}
        1 & : z < L_\mathrm{air} \\
        \varepsilon_\mathrm{YIG} & : x \in [ L_\mathrm{air},  L_\mathrm{air} + L_\mathrm{YIG}]  \\ 
        1 & : z \in [ L_\mathrm{air} + L_\mathrm{YIG}, L_\mathrm{cav}]  \\ 
        20 & : z \in [ L_\mathrm{cav}, L_\mathrm{cav} + L_\varepsilon]  \\ 
    \end{cases} \ ,
\end{align}
where $L_\mathrm{air}$ prescribes the placement of the YIG slab within the cavity, $L_\mathrm{YIG} = 50\,\mu\mathrm{m}$ is the length of the YIG slab, $L_\mathrm{cav} = 40.860\,\mathrm{mm}$ is the length of the cavity, and $L_\varepsilon = 2\,\mathrm{mm}$ is the thickness of the dielectric slab making up the right boundary of the cavity. In the left (right) column of Fig.~\ref{fig:1D_slab_in_cavity}, we consider a YIG slab placed at $L_\mathrm{air} = 10.753\,\mathrm{mm}$ ($L_\mathrm{air} = 11.674\,\mathrm{mm}$). Our choice of the YIG slab length $L_\mathrm{YIG}$ ensures that higher-order magnonic modes are negligible, while the cavity length $L_\mathrm{cav}$ is chosen so that the second-order cavity mode is tuned to the Kittel mode (\textit{i.e.}, two QNMs are considered in our calculations).

Figs.~\ref{fig:1D_slab_in_cavity}(a),(f), present the $H_0$-dependence of the QNM spectral positions, $\mathrm{Re}(\tilde{\omega}_n)$, evidencing level attraction (repulsion) for the left (right) configuration and signifying a transition from dissipative to coherent magnon-photon coupling. In Figs.~\ref{fig:1D_slab_in_cavity}(b),(g) we show the $H_0$-dependence of the QNM decay rates $\mathrm{Im}(\tilde{\omega}_n)$. The right configuration exhibits damping attraction, a typical signature of coherent coupling. For the left configuration, there is still a damping attraction, however, the rates do not cross. This is in contrast to the damping repulsion observed in the phenomenological model of dissipative magnon-photon coupling employed in Refs.~\cite{harder2018level, wang2020dissipative}, indicating an intermediate regime which the QNM theory is capable of describing from first principles. By further tuning the parameters to the transition point between these regimes of dissipative and coherent coupling, an exceptional point can be located, although we do not discuss it here. 
Next, Figs.~\ref{fig:1D_slab_in_cavity}(c),(h) present the QNM magnetic flux density profiles $|\tilde{B}^{(\mathrm{R})}_n(z)|$ for bias fields marked by the gray dashed lines in the panels above, namely $H_0 = 1.4844\,M_\mathrm{s}$ ($H_0 = 1.48725\,M_\mathrm{s}$) for the left (right) configuration (\textit{i.e.}, where the damping rates cross). The two QNMs have very similar mode profiles in both cases, particularly in the case of the right configuration where coherent coupling leads to the hybridization of the modes.
Note that the YIG slab resides at almost the same location relative to the cavity standing-wave mode in the two configurations, close to a node, which explains the sensitivity of the coupling mechanism to the slab position.

Figs.~\ref{fig:1D_slab_in_cavity}(d),(g) exhibit the magnetic Purcell factor for a QE position indicated by the grey dashed line in Figs.~\ref{fig:1D_slab_in_cavity}(c),(h), namely $z_\mathrm{QE} = L_\mathrm{cav}/2 = 20.430\,\mathrm{mm}$. For both configurations, the quantum QNM theory yields excellent agreement with the semi-analytical result obtained via the Green's function expansion in Eq.~\eqref{eqn:Greens_func_expansion}. In each case, we observe an apparent spectral splitting of the modes. For the left configuration, this splitting arises from the dissipative coupling of the modes, where the second mode contributes with opposite sign. This becomes especially clear upon a comparison with the predictions of the phenomenological QNM quantization scheme, which overestimates the Purcell factor and fails to reproduce the split-like lineshape. This is because the system operates in proximity to an exceptional point in its parameter space, where the two modal contributions separately diverge, while their sum remains finite~\cite{pick2017general}. Anomalous spontaneous emission spectra at exceptional points with a double peak structure have previously been reported for whispering-gallery-mode cavities~\cite{lu2022anomalous}. For illustration, we plot the two QNM contributions to the Green's function in Eq.~\eqref{eqn:Greens_func_expansion} individually, showing that one of them is completely negative. For the right configuration, the spectral splitting arises from the strong coherent coupling; in this case, the phenomenological QNM quantization correctly reproduces a spectral splitting, but not of the correct magnitude due to the additional dissipative coupling present in the system.

\begin{figure}
    \centering
    \includegraphics[width=0.8\linewidth]{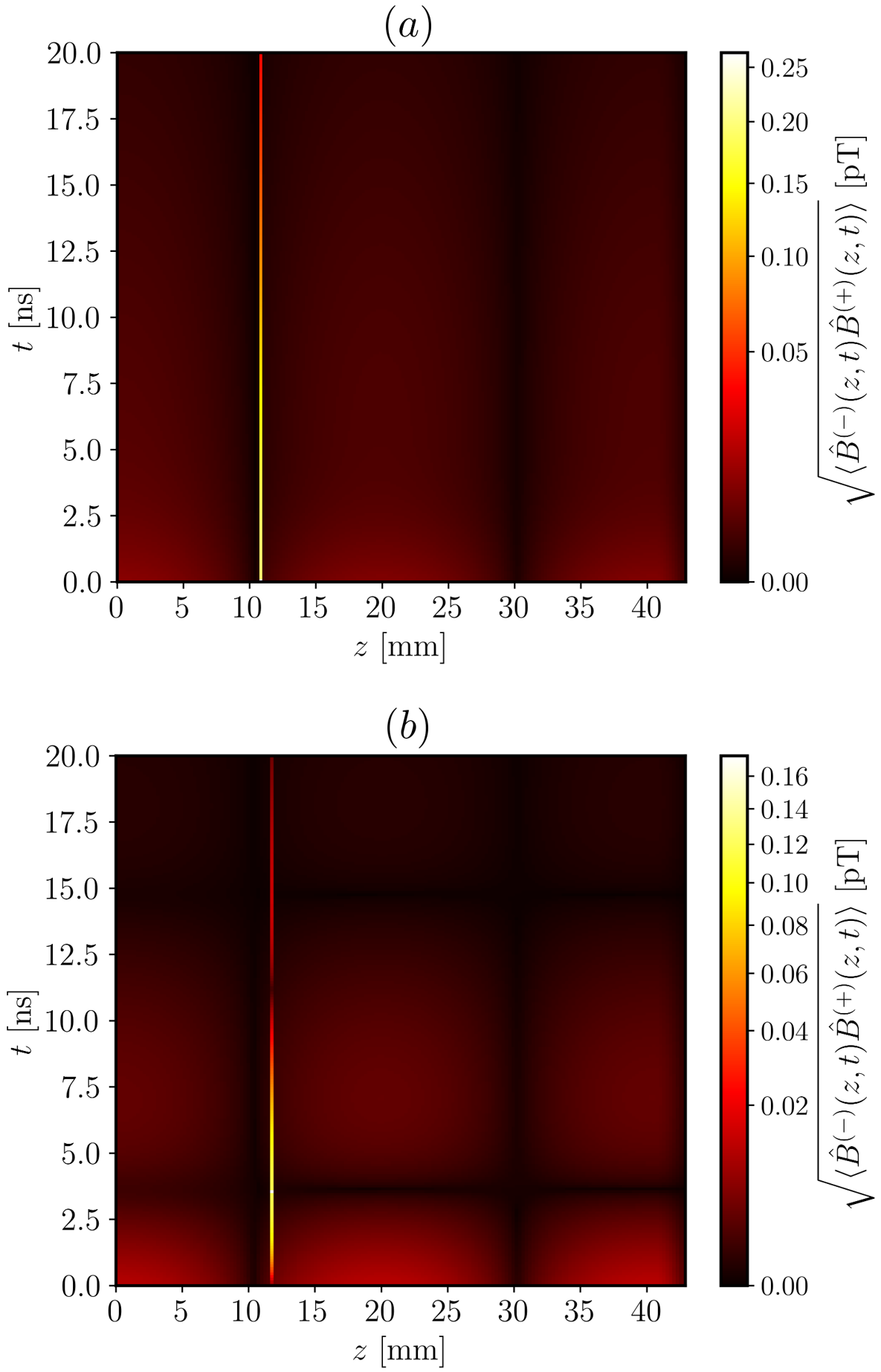}
    \caption{Spatiotemporally-resolved magnetic flux-density amplitude for the decay of the two-QNM Bell state $|\Psi^- \rangle = (|0 \rangle |1 \rangle - |1\rangle |0\rangle)/\sqrt{2}$ in the same cavity configuration of Fig.~\ref{fig:1D_slab_in_cavity} with (a) $L_{\mathrm{air}}=10.753\,\mathrm{mm}$ and (b) $L_{\mathrm{air}}=11.674\,\mathrm{mm}$.}
    \label{fig:1D_slab_in_cavity_spatiotemporal}
\end{figure}

Since our QNM quantization methodology provides a modal expansion of the electromagnetic field operators, it allows us to study the dynamics of the magnetic field in a spatially- and temporally-resolved manner, providing a powerful complement to the more common spatiospectral analyses in the literature. As a non-trivial example, Figs.~\ref{fig:1D_slab_in_cavity_spatiotemporal}(a),(b) present the spatiotemporal evolution of the magnetic flux-density amplitude $\langle \hat{B}^{(-)}(z,t) \hat{B}^{(+)}(z,t)\rangle^\frac{1}{2}$ for the decay of the initial two-QNM Bell state $|\Psi^- \rangle = (|0 \rangle |1 \rangle - |1\rangle |0\rangle)/\sqrt{2}$. For the left configuration, both the magnonic and photonic parts of the modes decay exponentially, whereas for the right configuration, coherent Rabi oscillations between the magnonic and photonic parts can be observed, enabled by the strong coherent magnon-photon coupling. Note that the QNMs are already the hybridized modes (\textit{i.e.}, superpositions of the magnonic and photonic parts), so an initial excitation of a single symmetrized mode would decay with much less pronounced Rabi oscillations; the Bell state is chosen precisely because it is a superposition in the two-mode QNM Fock basis. For the same reason, the Rabi oscillations are not simply a consequence of the off-diagonal coupling terms $\chi^\pm_{n n'}$, but would also appear without them.

These 1D examples demonstrate the versatility of our quantum QNM methodology, capturing both coherent and dissipative regimes of magnon-photon coupling from first principles, and revealing the limitations of purely phenomenological quantization approaches in describing cavity-QE interactions. The elucidation of the quantum field dynamics in both real space and real time constitutes a particularly valuable capability, and offers a truly four-dimensional characterization of quantum magnonic devices.

\subsection{\label{Subsec:3D_examples} 3D Spherical Systems}

\paragraph{Geometry and Qubit Couplings}

\begin{figure}
    \centering
    \includegraphics[width=1.0\linewidth]{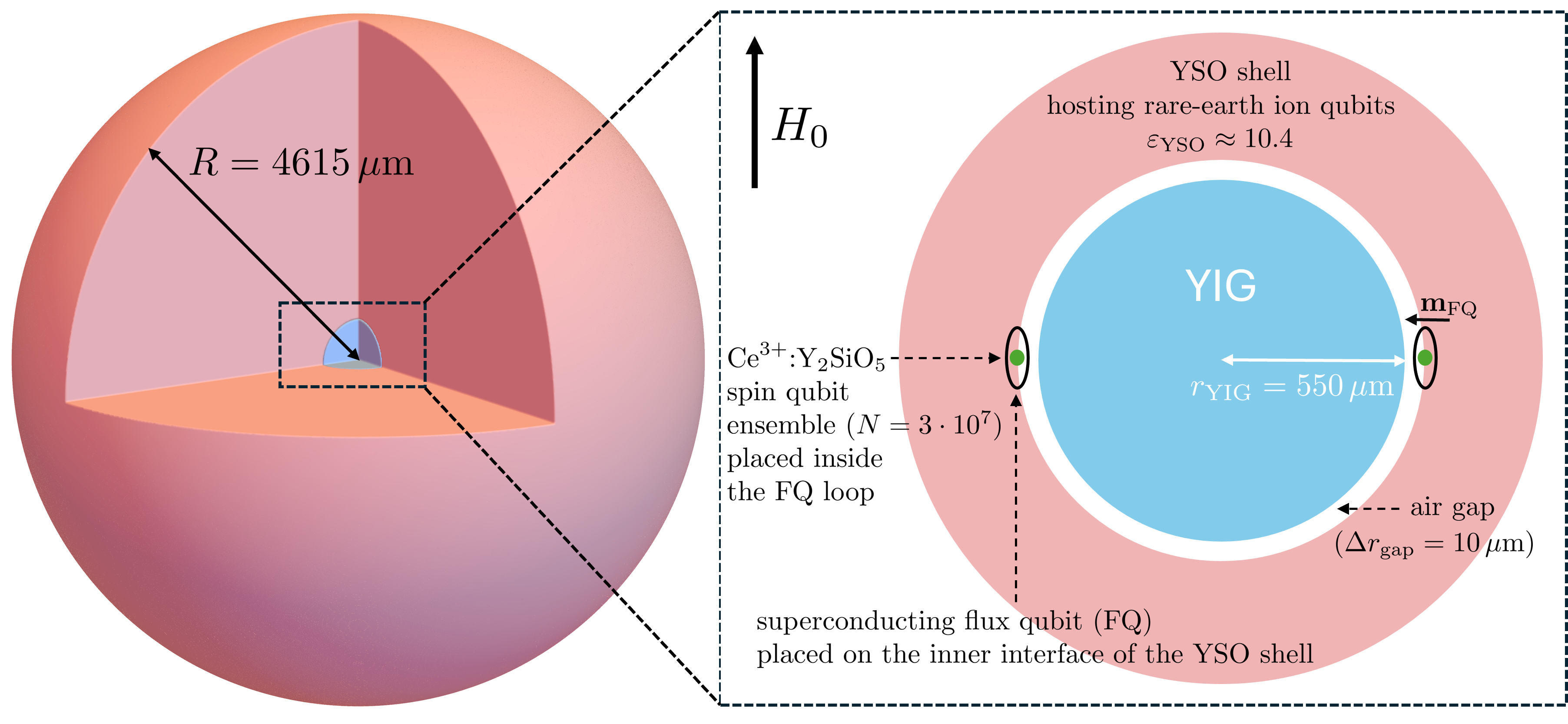}
    \caption{Schematic representation of the investigated spherically symmetric, core-shell, hybrid photonic-magnonic resonator device, comprising a YIG core and concentric YSO shell, together with embedded FQs and SQ ensembles. Relevant geometrical and physical parameters are indicated.}
    \label{fig:3D_schematic}
\end{figure}

In order to demonstrate the predictive capabilities of our quantum QNM theory in a 3D context, we explore a spherically symmetric, hybrid magnonic-photonic resonator and its coupling to FQs and SQs. The device is schematized in Fig.~\ref{fig:3D_schematic}. Specifically, we consider a spherical YIG core of radius $r_\mathrm{YIG} = 550\,\mu\mathrm{m}$ surrounded by a yttrium orthosilicate ($\mathrm{Y_{2}SiO_5 }$, YSO) shell of outer radius $R = 4615\,\mu\mathrm{m}$), with a $10\,\mu\mathrm{m}$ air gap between them. Two FQs (characterized by the persistent current $I_\mathrm{p} = 800\,\mathrm{nA}$ and loop radius $R_\mathrm{FQ}=10\,\mu\mathrm{m}$) reside on opposite sides of the YSO–air interface, with their loops oriented parallel to the direction of the bias field $\bm{H}_0 = H_0 \bm{e}_z$ ($H_0 \hat{=} H_0 - N_z M_\mathrm{s}$, $N_z = 1/3$) with $\bm{n}_\mathrm{FQ} = \bm{e}_y$. YSO can host rare-earth ions that function as SQs, and has previously been used as a substrate to couple SQs to a superconducting resonator~\cite{dold2019high}. For our numerical example, we choose an ensemble of $\mathrm{Ce}^{3+}$:$\mathrm{Y_2SiO_5}$ with $N=3\cdot10^{7}$ (corresponding to a doping of $\approx0.32\%$ of Ce1 sites in $1\,\mu\mathrm{m}^3$, assuming the Ce2 concentration to be negligible~\cite{laguta2017epr}), described as spin-$1/2$ systems with no zero field splitting and an effective $g$-tensor of $g = \mathrm{diag}(0.545, 1.609, 2.319)$~\cite{buryi2013electron}. The ensembles are assumed to be located inside the FQ loops, supporting a collective $\sqrt{N}$ scaling of the coupling constants. An important feature of our configuration is that, since the SQ ensembles reside inside the FQ loops, their direct interaction with the resonator modes is negligible and the dominant coupling is with the local FQ; the latter thus acts as a quantum interface to the resonator itself.

To describe the quantum dynamics arising from the QNM-FQ and FQ-SQ interactions, we now formulate the Hamiltonian of the system. Beginning with the FQ, its Hamiltonian in the persistent current basis is given by $\hat{H}_\mathrm{FQ} = -(\epsilon \tau_x + \Delta \tau_z)/2$, where we denote the FQ Pauli matrices by $\tau_i$ and $\epsilon = 2 I_\mathrm{p}( \Phi_\mathrm{ext} - \Phi_0/2)$, where $\Phi_\mathrm{ext}$ is the external magnetic flux through the FQ and $\Phi_0 = h/(2 e)$ is the magnetic flux quantum~\cite{dmitriev2021perspective,marcos2010coupling}. Decomposing the external magnetic flux into a static contribution, which maintains the qubit at the sweet spot ($\epsilon = 0$), and the resonator flux~\cite{kounalakis2022analog}
\begin{align}\label{eqn:FQ_flux_coupling}
    \hat{\Phi}_\mathrm{res} = \int_{A_\mathrm{FQ}}\mathrm{d}A\, \hat{\bm{B}}(\bm{r})\cdot \bm{n}_\mathrm{FQ} \approx A_\mathrm{FQ} (\hat{\bm{B}}(\bm{r}_\mathrm{FQ})\cdot \bm{n}_\mathrm{FQ} ) \ , 
\end{align}
leads to a Hamiltonian for the FQ coupled to the QNMs of the form $\hat{H}_\mathrm{FQ} = -\frac{\Delta}{2}\tau_x  + g^\mathrm{FQ}_n \hat{a}_n \tau_z + g_n^\mathrm{FQ*} \hat{a}^\dagger_n \tau_z$, where $g^\mathrm{FQ}_n = -I_\mathrm{p} A_\mathrm{FQ} (\hat{\bm{B}}(\bm{r}_\mathrm{FQ})\cdot \bm{n}_\mathrm{FQ})$. In principle, the integration in Eq.~\eqref{eqn:FQ_flux_coupling} should be performed over the FQ surface $A_\mathrm{FQ}$, with $\bm{n}_\mathrm{FQ}$ a unit normal vector to that surface.
The simple approximation used here is justified only if the symmetrized resonator field is nearly uniform over $A_\mathrm{FQ}$, which holds here as the magnonic resonator is much larger than the FQ and is placed in close proximity to it.

We model the interaction of a spin-1/2 SQ (neglecting zero-field splitting) with the resonator magnetic field via the Hamiltonian $\hat{H}_\mathrm{SQ} = {\mu_B}{B_0}(\bm{r}_\mathrm{SQ})g_z \sigma_z/2 + {\mu_B} \hat{\bm{B}}(\bm{r}_\mathrm{SQ})g \bm{\sigma}/2$, with the $g$-tensor $g = \mathrm{diag}(g_x,g_y,g_z)$ (diagonal in a system of principal axes) and the vector of Pauli matrices $\bm{\sigma} = (\sigma_x, \sigma_y, \sigma_z)^\mathrm{T}$. The first term describes the Zeeman splitting in the static bias field $B_0$, and the second term the coupling to the resonator. Approximating the FQ as a circular loop of radius $R_\mathrm{FQ}$ with the SQ at its center, the magnetic flux density generated by the persistent current is $\hat{\bm{B}}_\mathrm{FQ}(\bm{r}_\mathrm{SQ}) = {\mu_0} \hat{I}_\mathrm{P} \bm{n}_\mathrm{FQ} /{2 R_\mathrm{FQ}} = {\mu_0 I_\mathrm{P}}\bm{n}_\mathrm{FQ} \tau_z/{2 R_\mathrm{FQ}}$, and the FQ-SQ coupling Hamiltonian is given by $\hat{H}_\mathrm{FQ+SQ} = {\mu_\mathrm{B}} \hat{\bm{B}}_\mathrm{FQ}(\bm{r}_\mathrm{SQ}) g \bm{\sigma}/2$. Collecting these various contributions, applying the unitary transformation $\hat{U} = \exp(-\mathrm{i}\pi\tau_y/4)$ to transform the Hamiltonian into the FQ eigenbasis, and using the rotating wave approximation (\textit{i.e.}, keeping only excitation-conserving terms)~\cite{chang2025strong} leads to an effective Hamiltonian for the system comprising a FQ-SQ pair and the QNMs:
\begin{align}
    \hat{H} = +&\frac{\Delta}{2} \tau_z + \frac{\mu_0 B_0 g_z}{2} \sigma_z + \hbar\chi^+_{n n'} \hat{a}^\dagger_n \hat{a}_{n'} \notag \\
    + &g^\mathrm{FQ}_n \hat{a}_n \tau^+ + g^\mathrm{FQ*}_n \hat{a}^\dagger_n \tau^- + g^\mathrm{SQ}_n \hat{a}_n \sigma^+ + g^\mathrm{SQ*}_n \hat{a}^\dagger_n \sigma^- \notag \\
    +&g^\mathrm{SQ,z}_n \hat{a}_n \sigma_z + g^\mathrm{SQ,z*}_n \hat{a}^\dagger_n \sigma_z \notag \\
    + &g_{\mathrm{FQ-SQ}} \tau^+ \sigma^- + g^*_{\mathrm{FQ-SQ}} \tau^- \sigma^+ \notag \\
    + &g^z_{\mathrm{FQ-SQ}} \tau^+ \sigma_z + g^{z*}_{\mathrm{FQ-SQ}} \tau^- \sigma_z \ ,
\end{align}
with the transverse and longitudinal SQ-QNM coupling constants $g^\mathrm{SQ}_n = {\mu_\mathrm{B}}(g_x \tilde{B}^\mathrm{sym}_{n,x}(\bm{r}_\mathrm{SQ}) - \mathrm{i} g_y\tilde{B}^\mathrm{sym}_{n,y}(\bm{r}_\mathrm{SQ}) )/2$,  $g^\mathrm{SQ,z}_n = {\mu_\mathrm{B}} g_z \tilde{B}^\mathrm{sym}_{n,z}/2$, and the FQ-SQ coupling constants $g_\mathrm{FQ-SQ} = {\mu_\mathrm{B} \mu_0 I_\mathrm{p}} (g_x n_{\mathrm{FQ},x} + \mathrm{i}g_y n_{\mathrm{FQ},y})/{4 R_\mathrm{FQ}}$, $g^z_\mathrm{FQ-SQ} = {\mu_\mathrm{B} \mu_0 I_\mathrm{p}} g_z n_{\mathrm{FQ},z}/{4 R_\mathrm{FQ}}$. To account for qubit decay and pure dephasing effects, we supplement the quantum master equation with Lindblad dissipators of the form $\mathcal{D}[\hat{c}] = \hat{c}\rho \hat{c}^\dagger - 1/2\{\hat{c}^\dagger \hat{c},\rho \}$, where $\hat{c} = \sqrt{\gamma}\sigma^-$ and $\sqrt{\gamma_\varphi}\sigma_z$ respectively, with $\gamma = 1/T_1$ and $\gamma_\varphi = 1/T_2 - 1/(2 T_1)$, in which $T_{1}$ and $T_{2}$ are the corresponding relaxation times. The FQs are treated analogously.

\paragraph{Spatiospectral Characterization}

To calculate the QNMs and scattering cross-sections of the hybrid resonator described above, we employ the generalized Mie theory for anisotropic magnetic spheres~\cite{lin2004electromagnetic}, in conjunction with a semi-analytical transfer matrix approach for multilayer systems~\cite{zare2015magnetic}. Note that in the basis of vector spherical harmonics, the anisotropy of the permeability results in a mixing of transverse-electric (TE$_{lm}$) and transverse-magnetic (TM$_{lm}$) modes with different $l$; only the azimuthal index $m$ remains a good quantum number. 

\begin{figure}
    \centering
    \includegraphics[width=1.0\linewidth]{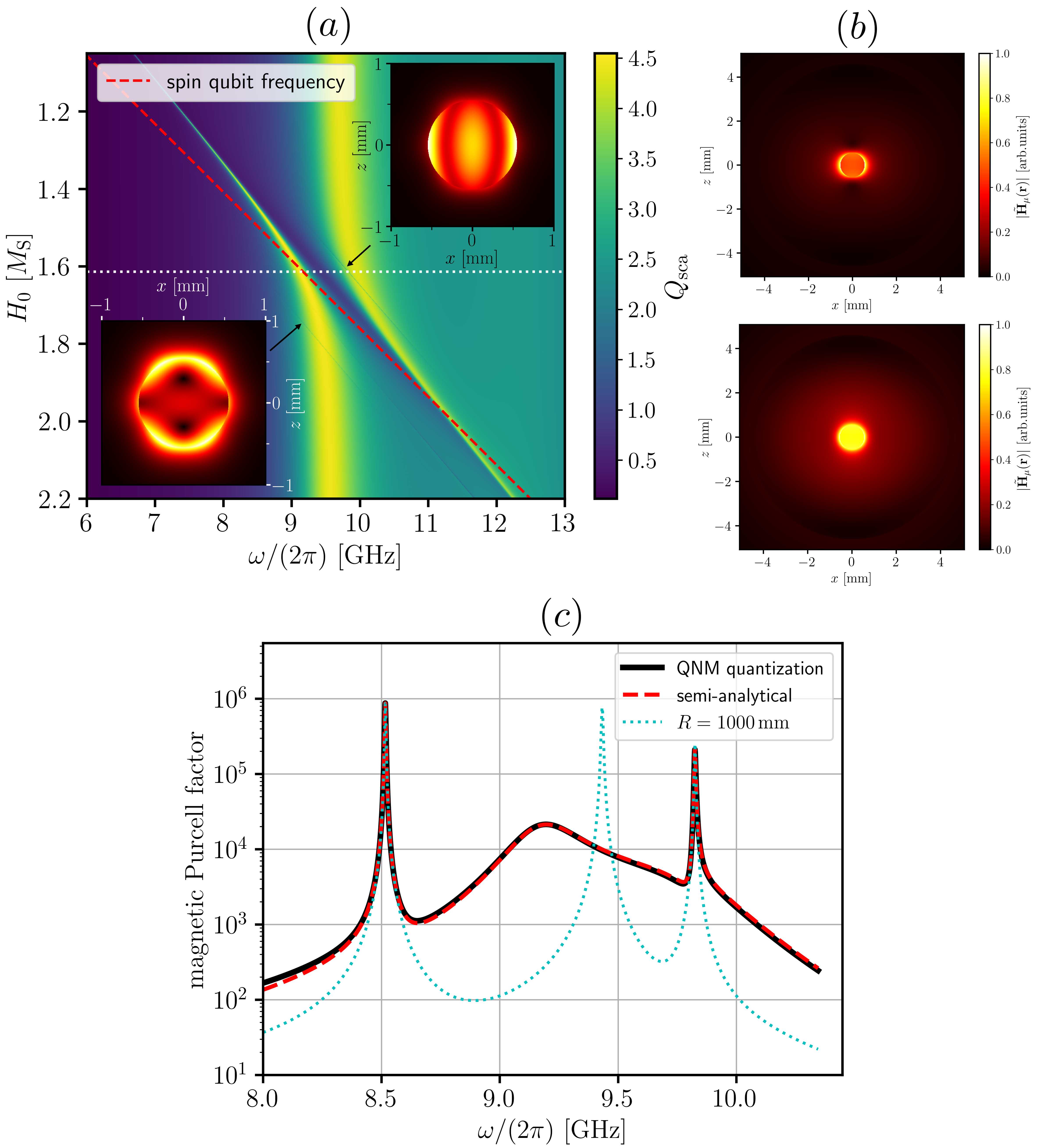}
    \caption{Spatiospectral characterization of the hybrid photonic-magnonic resonator schematized in Fig.~\ref{fig:3D_schematic}. (a) Scattering efficiency $Q_\mathrm{sca}$ as a function of bias field $H_0$ and frequency $\omega/(2\pi)$ for an incident, linearly polarized plane wave propagating along the positive $z$-direction. Additionally, the $H_0$-dependent SQ transition frequency and the magnetic field strength profiles $|\tilde{\bm{H}}^{(\mathrm{R})}_n(\bm{r})|$ of two higher-order magnonic modes with complex frequencies $\tilde{\omega}_n/(2\pi) = (8.5159 - 0.0026\mathrm{i})\,\mathrm{GHz}$ (lower left) and $(9.8244- 0.0035\mathrm{i})\,\mathrm{GHz}$ (upper right) for the bias field $H_0 = 1.614\,\mathrm{M}_\mathrm{s}$ (marked by the white dashed line) are shown. (b) Spatial magnetic field strength profiles $|\tilde{\bm{H}}^{(\mathrm{R})}_n|$ of the two hybridized magnon-photon modes with complex frequencies $\tilde{\omega}_n/(2\pi) = (9.1711- 0.1603\mathrm{i})\,\mathrm{GHz}$ (upper panel) and $(9.7941- 0.3476\mathrm{i})\,\mathrm{GHz}$ (lower panel). (c) Frequency-dependent magnetic Purcell factors of the FQs calculated from the quantum QNM theory incorporating the four QNMs from (a) and (b), compared with the semi-analytical result obtained from the QNM expansion of the Green's function. Additionally, the semi-analytical result for the same hybrid resonator device, but with a YSO shell of radius $R = 1000\,\mu\mathrm{m}$, is shown.}
    \label{fig:3D_synchronization}
\end{figure}

Fig.~\ref{fig:3D_synchronization}(a) shows the $H_0$- and $\omega$-dependent scattering efficiency $Q_\mathrm{sca}$, assuming an incident electromagnetic plane wave propagating along the $z$-direction. We observe a marked anticrossing of two modes; the $H_0$-independent branch corresponds to the lowest-order $\mathrm{TE}_{1 m}$ photonic mode of the YSO shell, while the $H_0$-dependent one corresponds to the magnonic Kittel mode of the YIG sphere, predominantly $\mathrm{TE}_{1 1}$ in character. Additionally, we plotted the $H_0$-dependent Zeeman transition frequency of the SQs, determined by the $g$-tensor component $g_z = 2.319$, and marked the bias field $H_0 = 1.614$ chosen for the preceding calculations where the SQs are tuned to the lower hybridized mode.
Apart from the Kittel and photonic modes, three additional $H_0$-dependent features can be observed in the scattering efficiency map, manifesting higher-order magnonic QNMs. The middle one progressively weakens with increasing bias field and becomes negligible for $H_0 = 1.614$; the other two QNMs, indicated by black arrows and corresponding spatial field profiles $|\tilde{\bm{H}}^{(\mathrm{R})}_n(\bm{r})|$ in Fig.~\ref{fig:3D_synchronization}(a), have a core-localized nature with predominantly $(\mathrm{TE}_{1 1} + \mathrm{TM}_{2 1} + \mathrm{TE}_{3 1})$ character (\textit{i.e.}, mixed spherical multipoles), barely coupling to the photonic mode due to their inhomogeneous radial profile in the $\mathrm{TE}_{11}$-component.
Fig.~\ref{fig:3D_synchronization}(b) exhibits the magnetic field strength profiles pertaining to the two hybridized magnon-photon modes, with complex frequencies $\tilde{\omega}_n/(2\pi) = (9.1711- 0.1603\mathrm{i})\,\mathrm{GHz}$ (upper plot) and $(9.7941- 0.3476\mathrm{i})\,\mathrm{GHz}$ (lower plot) for $H_0 = 1.614$. The lower-frequency mode exhibits a stronger magnonic character and is mostly homogeneous within the YIG sphere, reflecting its origin in the Kittel mode. Near the FMR, one of the eigenvalues of the permeability becomes negative ($\mathrm{Re}(\mu_{xx}(\omega) + \kappa(\omega)) < 0$), giving rise to a surface-localized field profile analogous to a plasmonic resonance, yet not a true surface plasmon of the kind observed in metallic nanostructures. The higher-frequency mode possesses more photonic character, but also shows a hotspot inside the YIG sphere rather than surface field localization.

Similar to \S~{\ref{Subsec:1D_examples}}, we further characterize our hybrid resonator via a calculation of the magnetic Purcell factor, plotted for the FQs in Fig.~\ref{fig:3D_synchronization}(c). Our results, obtained from a four-mode QNM expansion in the full quantum treatment, stand in good agreement with the semi-analytical prediction via the Green's function. Notably, the hybridized magnon-photon modes are much broader than the higher-order magnonic modes and have much lower Purcell factors ($\sim 0.88\times10^6$ and $0.21\times10^6$ for the higher-order modes), attributable to the hybridization with the photonic mode. For comparison, the same (semi-analytical) calculation with a thinner YSO shell ($R=1000\,\mu\mathrm{m}$), which supports no photonic mode in the considered frequency range, shows that the unhybridized Kittel mode has a much larger Purcell factor. Interestingly however, the lower-frequency, higher-order magnon mode offers a slightly larger Purcell factor and a higher quality factor, so that if there are the relevant figures of merit in given device application, the higher-order magnonic modes might actually be more suitable.

\paragraph{Magnon-mediated Qubit Synchronization and Entanglement Dynamics}

\begin{figure}
    \centering
    \includegraphics[width=1.0\linewidth]{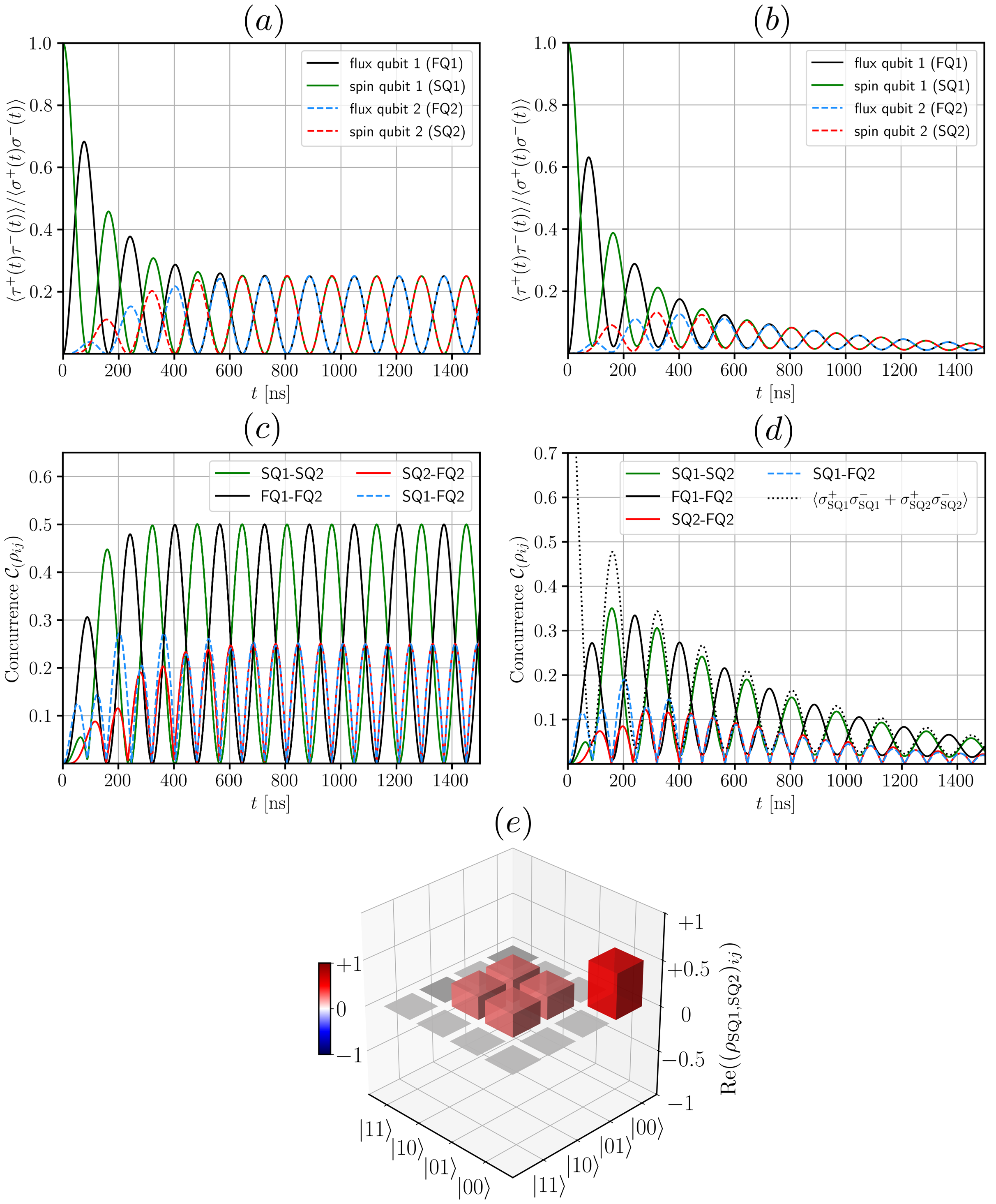}
    \caption{Qubit dynamics in the hybrid photonic-magnonic resonator schematized in Fig.~\ref{fig:3D_schematic}. (a),(b) SQ and FQ dynamics for one initially excited SQ in the case of no qubit damping or dephasing (a) and for relaxation times $T_{1} = 1.47\,\mu\mathrm{s}$ and $T_{2} = 1.45\,\mu\mathrm{s}$ (b). (c),(d) Time-dependent concurrences for two-qubit subsystems $\mathcal{C}(\rho_{A B})$ in the case of no qubit damping or dephasing (c), and for damped qubits (d). (e) Real part of the two-SQ reduced density-matrix elements $\mathrm{Re}((\rho_{\mathrm{SQ1},\mathrm{SQ2}})_{i j})$ in the case of no qubit damping or dephasing, at times of maximum concurrence $\mathcal{C}(\rho_{\mathrm{SQ1},\mathrm{SQ2}})$.}
    \label{fig:3D_synchronization_temporal}
\end{figure}

Turning to the temporal evolution of the system, Fig.~\ref{fig:3D_synchronization_temporal}(a) exhibits the qubit dynamics assuming one SQ initially excited and the FQ transition tuned to the SQs. Each SQ undergoes strong coupling with its local FQ and Rabi oscillations can be observed. Through their interaction with a common resonator, excitation is transferred to the opposite FQ-SQ pair, where similar Rabi oscillations develop. After about four periods, both the SQs and the FQs are synchronized in their oscillations. This phenomenon is in accordance with noise-induced quantum synchronization in qubit chains~\cite{schmolke2022noise,tao2025noise}, where the application of noise to the transition frequency of the middle qubit in the chain creates a decoherence-free subspace (\textit{i.e.}, a decoupling from the environment) that sustains the qubit oscillations. In our case, the middle qubit is replaced by the resonator, the noise is provided by the lossy mode, and we utilize different types of qubits (FQs and SQs). Even though the mode exhibits loss rather than just noise and the qubits couple to it, the amount of excitation present in the system does not further decrease once the qubits are synchronized (\textit{i.e.}, once the pair is synchronized, the lossy mode does not drain excitation).

Such behavior is not expected in the presence of qubit relaxation effects however. In Fig.~\ref{fig:3D_synchronization_temporal}(b), we present the qubit dynamics in which both the SQs and FQs possess the relatively short relaxation times $T_{1} = 1.47\,\mu\mathrm{s}$ and $T_{2} = 1.45\,\mu\mathrm{s}$, taken from recently reported experimental results for zero-flux-biased FQs~\cite{kim2024superconducting}. Note that rare-earth SQs usually have better $T_{1,2}$ values, but since the decoherence of the FQs is the limiting factor here, we simply choose the same values for the SQs. The synchronization and the transient time to reach the quasi-steady state are not affected by the presence of qubit damping or dephasing, the excitation present in the system only decays over time according to the timescale $T_1$; such predictions also remain valid for other combinations of SQ/FQ $T_{1,2}$ values.

Deeper insight into the quantum dynamics of the multiqubit-resonator system can be acquired by exploring the time-dependent entanglement of the various two-qubit subsystems. To this end, we quantify the degree of entanglement between qubits $A$ and $B$ via the concurrence $\mathcal{C}(\rho_{A B})$~\cite{horodecki2009entangl}, shown in Fig.~\ref{fig:3D_synchronization_temporal}(c) for the idealized case in which there is no qubit damping or dephasing. For the SQ-SQ and FQ-FQ subsystems, the concurrences follow the excitation oscillations with maxima of $\mathrm{max}(\mathcal{C}(\rho_{\mathrm{SQ1},\mathrm{SQ2}}(t))) \approx \mathrm{max}(\mathcal{C}(\rho_{\mathrm{FQ1},\mathrm{FQ2}}(t))) \approx 0.5$, implying that the two SQs or FQs possess the maximum entanglement allowed by their reduced excitation probabilities in Fig.~\ref{fig:3D_synchronization_temporal}(a). The FQ-SQ concurrences oscillate at double that frequency, with a phase shift, and have $\mathrm{max}(\mathcal{C}(\rho_{\mathrm{SQ1/2},\mathrm{FQ1/2}}(t))) \approx \mathrm{max}(\mathcal{C}(\rho_{\mathrm{SQ1/2},\mathrm{FQ2/1}}(t))) \approx 0.25$ at the times where FQs and SQs have equal excitation probability. Each SQ is thus entangled with the FQ on both the same side and the opposite side. The real parts of the two-SQ reduced density-matrix elements $\mathrm{Re}((\rho_{\mathrm{SQ1},\mathrm{SQ2}})_{i j})$ are shown in Fig.~\ref{fig:3D_synchronization_temporal}(e) for a temporal instant of maximum concurrence, resembling the state $\rho= p_{1} | \Psi^- \rangle \langle \Psi^- | + p_{2} | 0 0 \rangle \langle 00 | + p_{3} | \Psi^+ \rangle \langle \Psi^+ | + p_4 | 0 0 \rangle \langle 00 |$, in which $p_{2} = p_{3 } = 1/2$  and $p_{1} = p_{4 } = 0$, with a fidelity of $F \approx 99.94\%$ (adopting the convention used in QuTiP~\cite{lambert2024qutip}). The same results is found for $\rho_{\mathrm{FQ1},\mathrm{FQ2}}$. Notably, this is a different state from that of the five-qubit chain investigated in Ref.~\cite{tao2025noise}, where $p_1 = 1/2$, $p_2 = 1/3$, $p_3 = 1/6$ and $p_4 = 0$, though in both cases, they are maximally entangled mixed states~\cite{ishizaka2000maximally}. At times where the FQs and SQs have equal excitation probability, the FQ-SQ subsystems are in the states $\rho_{\mathrm{FQ1},\mathrm{SQ1}} = \rho_{\mathrm{FQ2},\mathrm{SQ2}} = 0.25 |\Psi^{-\mathrm{i}} \rangle \langle \Psi^{-\mathrm{i}} | + 0.75 |00 \rangle \langle 00|$ and $\rho_{\mathrm{FQ1},\mathrm{SQ2}} = \rho_{\mathrm{FQ2},\mathrm{SQ1}} = 0.25 |\Psi^{+\mathrm{i}} \rangle \langle \Psi^{+\mathrm{i}} | + 0.75 |00 \rangle \langle 00|$, with $|\Psi^{\pm\mathrm{i}} \rangle = ( |0 1\rangle \pm \mathrm{i |1 0 \rangle})/\sqrt{2}$ (\textit{i.e.}, also maximally entangled mixed states). 

The impact of decay and pure dephasing are evidenced in Fig.~\ref{fig:3D_synchronization_temporal}(d), which presents the same concurrences in the case of qubits with relaxation times of $T_{1} = 1.47\,\mu\mathrm{s}$ and $T_{2} = 1.45\,\mu\mathrm{s}$. The concurrences follow the same qualitative behavior, but with maxima that follow a decaying envelope as they are bounded by the reduced excitation probabilities. Additionally, we plot the combined SQ excitation probabilities $p_\mathrm{SQs} = \langle \sigma^+_\mathrm{SQ1} \sigma^-_\mathrm{SQ1} + \sigma^+_\mathrm{SQ2} \sigma^-_\mathrm{SQ2} \rangle$, showing that the concurrence $\mathcal{C}(\rho_{\mathrm{SQ1},\mathrm{SQ2}})$ does not fully reach its upper bound at the times of its local maxima (\textit{i.e}, the fidelity of the state is reduced). However, for $t > 400\,\mathrm{ns}$, fidelity values of $>99.3\%$ are still achieved for the SQ-SQ subsystem and $\rho(t) = p_\mathrm{SQs}(t) |\Psi^+ \rangle \langle\Psi^+ | + (1 -p_\mathrm{SQs}(t) ) |00 \rangle \langle 00 |$.

\begin{figure}
    \centering
    \includegraphics[width=1.0\linewidth]{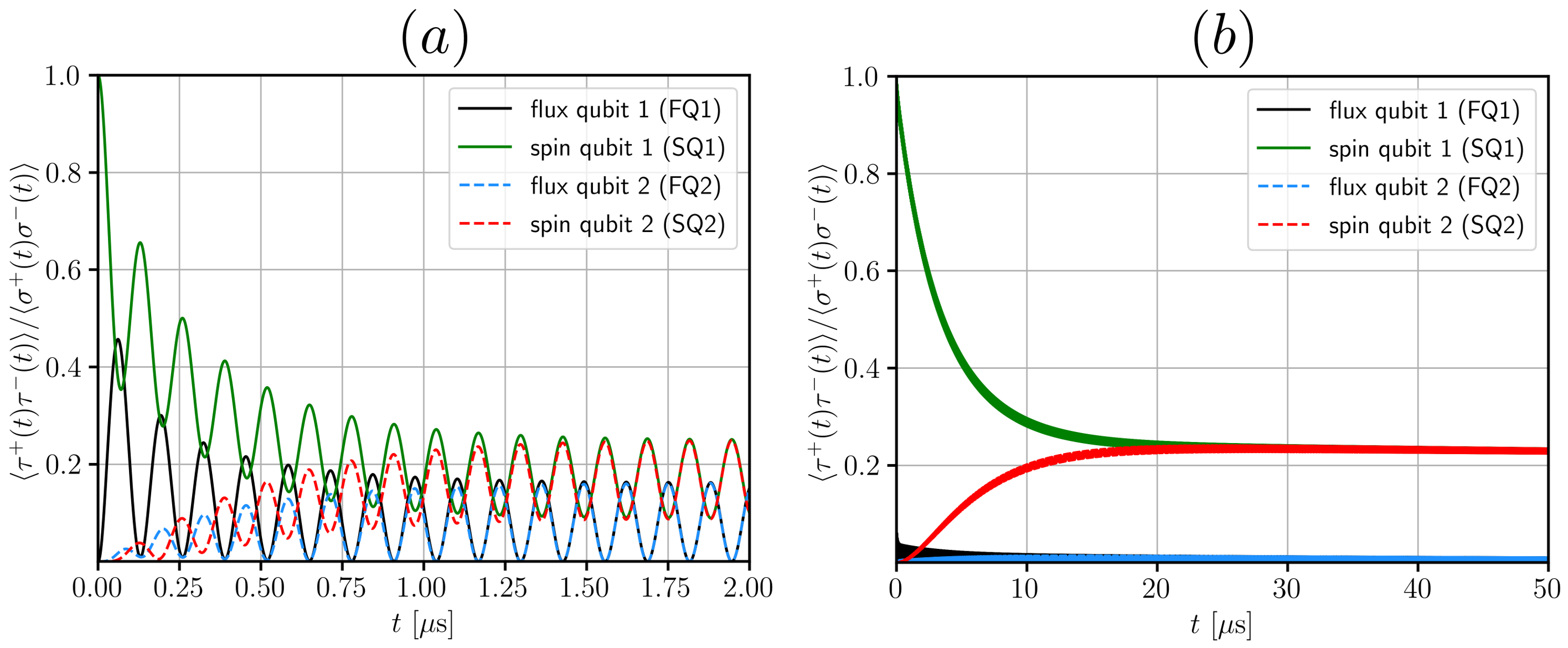}
    \caption{Numerical results for detuned qubit dynamics, assuming one initially excited SQ. (a) FQs are detuned to $\omega_\mathrm{FQ} = 1.0005 \omega_\mathrm{SQ}$ ($\omega_\mathrm{FQ} - \omega_\mathrm{SQ} = 1.479 g_{\mathrm{FQ-SQ}}$) in the absence of decay and pure dephasing. (b) FQs are detuned to $\omega_\mathrm{FQ} = 1.0025 \omega_\mathrm{SQ}$ ($\omega_\mathrm{FQ} - \omega_\mathrm{SQ} = 7.393 g_{\mathrm{FQ-SQ}}$), in the presence of decay and pure dephasing with $T_{1} = 40\,\mu\mathrm{s}$ and $T_{2} = 85\,\mu\mathrm{s}$ (FQs), and $T_{1} = T_{2} = 3\,\mathrm{ms}$ (SQs). }
    \label{fig:3D_detuning}
\end{figure}

Whilst not discussed in further detail here, two general conclusions can be made from our simulated quantum dynamics.
First, the loss rate of the resonator mode affects the transition time required for the system to attain a quasistationary synchronized state and an optimum appears to exist (\textit{i.e.}, for too large or too small loss rates, the transient time increases). However, the amount of excitation present in the system in its quasistationary state does not depend on the loss rate of the mode. The same is also true for the FQ-resonator coupling. Second, detuning the FQs from the SQs leaves the maximally entangled mixed state of the SQ subsystem intact at the times of maximum concurrence, but reduces the fraction of excitation that is transferred to the FQs. This is visible for the undamped case with $\omega_\mathrm{FQ} - \omega_\mathrm{SQ}  = 1.479 g_{\mathrm{FQ-SQ}}$ in Fig.~\ref{fig:3D_detuning}(a); the FQs reach a maximally entangled mixed state of lower excitation, and the transient time for synchronization is lengthened. Increasing the detuning further suppresses the transfer until the SQs retain nearly all of the excitation, so that the system settles into a long-lived, weakly oscillating state without continuous driving. This regime is illustrated in Fig.~\ref{fig:3D_detuning}(b) for damped qubits ($T_{1} = 40\,\mu\mathrm{s}$ and $T_{2} = 85\,\mu\mathrm{s}$ for the FQs~\cite{yan2016flux} and $T_{1} = T_{2} =  3\,\mathrm{ms}$ for the SQs) at $\omega_\mathrm{FQ} - \omega_\mathrm{SQ}  = 7.393 g_{\mathrm{FQ-SQ}}$; the longer coherence times were chosen so that the extended transient remains observable, given that the transient time also grows with the detuning. The SQ subsystem reaches the maximally entangled mixed state with a fidelity of approximately $99.85\,\%$. Note that for rare-earth ions in YSO, the superhyperfine interaction with neighboring nuclear spins could limit the achievable coherence times in practice, but nevertheless, optimized host crystals have demonstrated $T_{2}$ values up to approximately $23\,\mathrm{ms}$~\cite{le2021twenty}.

\paragraph{Finite-temperature Effects}

\begin{figure}
    \centering
    \includegraphics[width=1.0\linewidth]{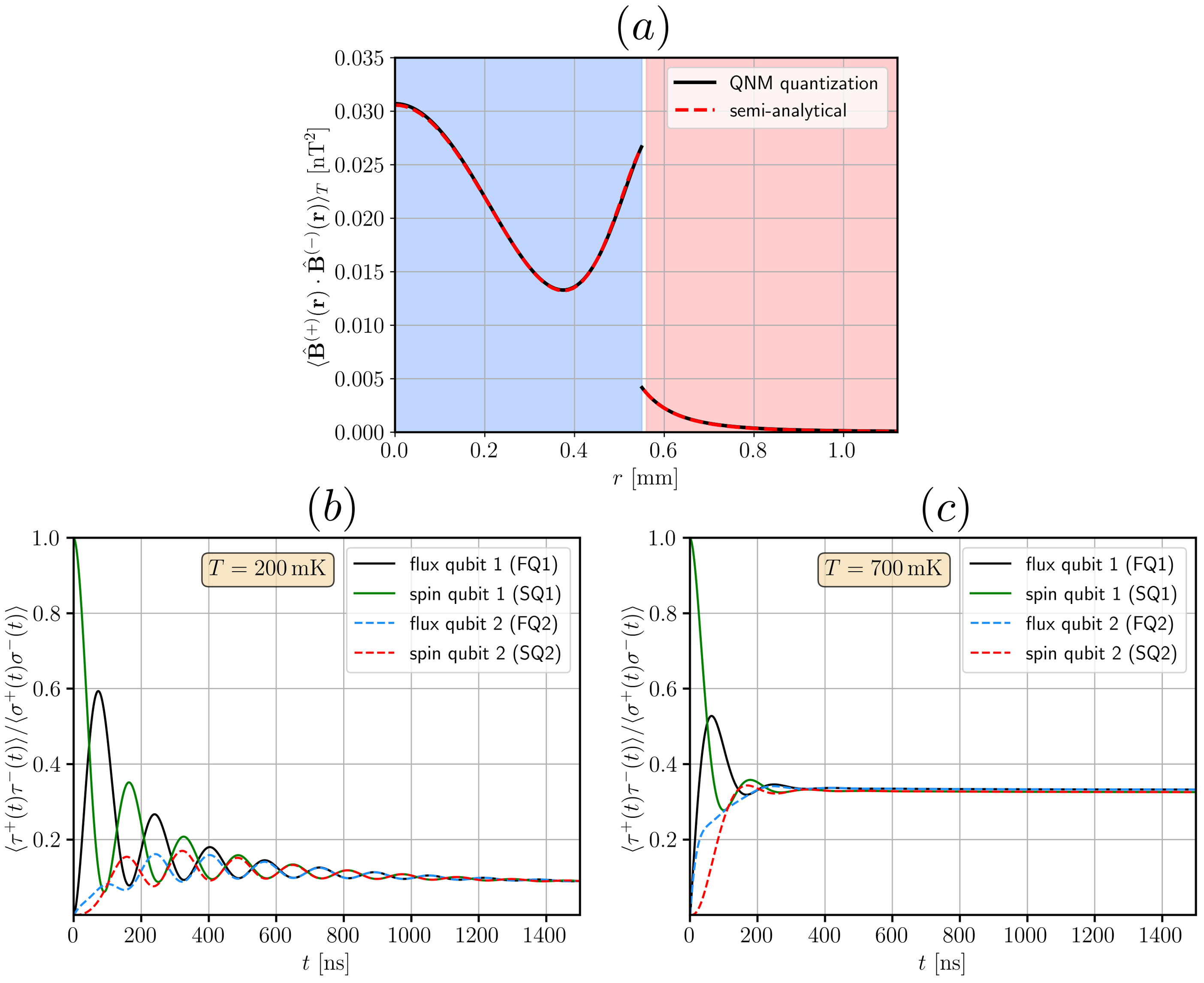}
    \caption{Numerical results for the hybrid photonic-magnonic resonator schematized in Fig.~\ref{fig:3D_schematic} and characterized in Fig.~\ref{fig:3D_synchronization} at finite temperatures. (a) Comparison of spatially resolved magnetic-field thermal correlation functions at $T=450\,\mathrm{mK}$, calculated from the quantum QNM theory and semi-analytical approach. (b),(c) Qubit dynamics of one initially excited SQ at $T=200\,\mathrm{mK}$ (b) and $T=700\,\mathrm{mK}$ (c), assuming $T_{1} = 1.47\,\mu\mathrm{s}$ and $T_{2} = 1.45\,\mu\mathrm{s}$ for all qubits.}
    \label{fig:3D_finite_temp}
\end{figure}

Finally, we demonstrate the capability of our quantum QNM approach to capture finite-temperature effects in the hybrid magnonic-photonic resonator device of Fig.~{\ref{fig:3D_schematic}}, following the formalism in \S~\ref{subsec:finite_temp}. In Fig.~\ref{fig:3D_finite_temp}(a), we present the radial dependence of the magnetic-field thermal correlation function $\langle \hat{\bm{B}}^{(+)}(\bm{r}) \hat{\bm{B}}^{(-)}(\bm{r}) \rangle_T$ (see Eq.~\eqref{eqn:E_field_thermal_corr}) in the $xy$ plane, for $T = 450\,\mathrm{mK}$, where the frequency integrals were computed for $\omega/(2\pi) \in [0\,\mathrm{GHz}, 10.5\,\mathrm{GHz}]$. We compare the result from the QNM quantization with the semi-analytical one obtained from the imaginary part of the Green's function, evidencing excellent agreement. Fig.~\ref{fig:3D_finite_temp}(b) shows the qubit dynamics ($T_{1} = 1.47\,\mu\mathrm{s}$ and $T_{2} = 1.45\,\mu\mathrm{s}$) in an environment with $T=200\,\mathrm{mK}$. The synchronization itself is not affected, however there is additional dephasing and the system relaxes towards a non-oscillating thermal steady state. With increasing temperature, the system relaxes more rapidly towards its steady state, as shown in Fig.~\ref{fig:3D_finite_temp}(c) for $T=700\,\mathrm{mK}$. As only barely half a synchronized oscillation can be observed, the critical temperature of this device (\textit{i.e.}, the temperature above which the desired functionality can no longer be observed) is approximately $T_\mathrm{crit} = 700\,\mathrm{mK}$.

\section{Conclusion and Outlook}

In this work, we have introduced a finite-temperature quantum QNM theory for 3D, dissipative and dispersive, linear, cavity magnonic systems and their interaction with spin and superconducting qubits. At the level of the classical theory, we have generalized the auxiliary-field eigenvalue methodology for the computation of the classical QNMs of gyrotropic magnetic resonators to include both the exchange interaction and anisotropy effects, and discussed QNM expansions for the Green's function and the calculation of other important quantities. Invoking the theory of MQED for non-reciprocal media, we have then generalized our PML-based QNM quantization approach of Ref.~\cite{meschede2026quantum} for lossy magnonic resonators in arbitrary (linear) dielectric environments, enabling the prediction of all magnon-photon couplings and hybridizations from first principles, revealing the quantitative limitations of purely phenomenological quantization protocols, and offering unprecedented insight into the quantum dynamics of magnon-qubit interactions in hybrid quantum architectures. Moreover, we have taken into account finite-temperature effects, which play a more significant role in cavity magnonic systems, relative to dielectric or plasmonic ones, by virtue of their lower (RF and MW) excitation frequencies.

The predictive capabilities of our methodology have been demonstrated via application to a 1D, YIG slab resonator, capturing the accumulation of higher-order magnonic modes at the pole of the effective permeability and their interference with the Kittel mode. When the same slab is placed inside a photonic cavity, our method has enabled an essentially exact prediction of the continuous tuning, via sub-millimeter geometrical changes, from dissipative to coherent magnon–photon coupling, with the associated Purcell spectra and spatiotemporally-resolved field dynamics ascertained. Indeed, even in the proximity of an exceptional point, where two modal contributions to the Green's function themselves diverge and a phenomenological quantization scheme fails completely, the quantum QNM method is able to correctly describe the nature of the system.
Motivated by recent progress in directly interfacing SQs or FQs and magnonic resonators~\cite{gunzler2025spin, kim2024superconducting, zhu2011coherent, twamley2010superconducting, marcos2010coupling,chang2025strong}, we have furthermore demonstrated the application of our quantum QNM theory to a 3D YIG-YSO hybrid resonator coupled to such qubits, where the photonic component (YSO shell) is similar in size to the YIG sphere, and consequently perturbative treatments of the magnon-photon coupling may break down~\cite{macedo2020electromagnetic}. Our calculations not only yield the hybrid modes and qubit couplings, but show how the tuning of losses enabled by the hybrid magnonic-photonic nature of the construct can enable functionalities like magnon-mediated synchronization and entanglement of spatially separate qubits, as well as attaining almost-steady states in the absence of continuous pumping, including at finite temperature and in the presence of realistic qubit decoherence effects. Ultimately, our numerical examples exhibit how the quantum QNM approach can be utilized as a quantitative basis for the design, optimization and analysis of quantum cavity magnonic devices as a functional elements of hybrid quantum processors operating under open-system and thermal conditions.

Our work admits a variety of natural extensions and seeds important further developments in the theoretical treatment of quantum cavity magnonics. For example, incorporating the exchange interaction
in numerical QNM solvers, such as the MAN programs~\cite{wu2021qnmnonreciprocal_resonators,wu2021nanoscale,wu2023man}, and quantifying its impact on the behavior and performance of quantum magnonic devices will be crucial as such technologies are driven down to ever smaller length scales.
Furthermore, it is of interest to explore nonlinear effects that can, for example, lead to the suppression of magnon-photon hybridization~\cite{lee2023nonlinear}, while another promising route is the inclusion of time-periodic fields that opens the route to Floquet cavity electromagnonics, with a plethora of intriguing phenomena like the emergence of sidebands, magnonic Autler-Townes splittings, Floquet ultrastrong coupling, and the coherent control of magnon-photon coupling~\cite{xu2020floquet,aguiar2026perturbative,hosseini2026time}. Recent progress in classical QNM theory~\cite{vial2025quasinormal,sun2026quasinormal} and MQED~\cite{horsley2025macroscopic} for Floquet systems have already established important foundations, so that such developments may lie within reach.

\begin{acknowledgments}
This publication has emanated from research conducted with the financial support of Research Ireland, under Grants Number 18/RP/6236 and 23/RC/12197 at Rinn Quantum.
\end{acknowledgments}

\section*{Conflict of Interest}
The authors have no conflicts to disclose.

\section*{Data Availability}
The data that support the findings of this study are available from the corresponding authors upon reasonable request.

\bibliography{aipsamp}

\end{document}